\documentclass[aps,pre,superscriptaddress,balancelastpage,nofootinbib]{revtex4-2}

\usepackage[colorlinks,bookmarks=false,citecolor=blue,linkcolor=blue,urlcolor=blue]{hyperref}
\usepackage[all]{hypcap}   % let hyperlinks correctly point to figures rather than their captions;

\usepackage{amsmath,amssymb}
\usepackage{graphicx}
\usepackage{bbold}

\usepackage{verbatim}
\usepackage{color}

\usepackage{placeins}    % for FloatBarrier
\usepackage{flafter}     % Bilder immer nach \figure Befehl
\usepackage{color}

\usepackage[normalem]{ulem}

\usepackage[capitalise]{cleveref}
\AddToHook{cmd/appendix/before}{%
    \crefalias{section}{appendix}%
    \crefalias{subsection}{appendix}
}
\crefname{section}{Sec.}{Secs.}

\newcommand{\ue}{\text{e}}
\newcommand{\ui}{\text{i}}
\newcommand{\ud}{\text{d}}

\newcommand{\Imag}{\ensuremath{\mathrm{Im}}}
\newcommand{\Real}{\ensuremath{\mathrm{Re}}}

\newcommand{\vecr}{\ensuremath{\boldsymbol{r}}}
\newcommand{\psiR}{\ensuremath{\psi^\mathrm{R}}}
\newcommand{\psiL}{\ensuremath{\psi^\mathrm{L}}}
\newcommand{\nref}{\ensuremath{n_{\mathrm{r}}}}
\newcommand{\nrefsq}{\ensuremath{n_{\mathrm{r}}^2}}

\newcommand{\BVec}{\ensuremath{\boldsymbol{B}}}
\newcommand{\GVec}{\ensuremath{\boldsymbol{G}}}
\newcommand{\JVec}{\ensuremath{\boldsymbol{J}}}
\newcommand{\MVec}{\ensuremath{\boldsymbol{M}}}
\newcommand{\MVecTilde}{\ensuremath{\boldsymbol{\tilde{M}}}}
\newcommand{\RVec}{\ensuremath{\boldsymbol{R}}}
\newcommand{\RVecTilde}{\ensuremath{\boldsymbol{\tilde{R}}}}
\newcommand{\nuVec}{\ensuremath{\boldsymbol{\nu}}}
\newcommand{\tVec}{\ensuremath{\boldsymbol{t}}}
\newcommand{\dV}{\ensuremath{\ud \mathcal{V}}}
\newcommand{\dS}{\ensuremath{\ud \mathcal{S}}}
\newcommand{\EVec}{\ensuremath{\boldsymbol{E}}}
\newcommand{\HVec}{\ensuremath{\boldsymbol{H}}}
\newcommand{\murel}{\ensuremath{\mu_{\mathrm{r}}}}
\newcommand{\epsrel}{\ensuremath{\varepsilon_{\mathrm{r}}}}
\newcommand{\mutens}{\ensuremath{\boldsymbol{\mu}}}
\newcommand{\epstens}{\ensuremath{\boldsymbol{\varepsilon}}}

\begin{document}

\title{Norm of resonance states in \\
quantum scattering and electromagnetic systems}

\author{Florian Lorenz}
\affiliation{TU Dresden,
 Institute of Theoretical Physics and Center for Dynamics,
 01062 Dresden, Germany}

\author{Jan Möseritz-Schmidt}
\affiliation{TU Dresden,
 Institute of Theoretical Physics and Center for Dynamics,
 01062 Dresden, Germany}

\author{Roland Ketzmerick}
\affiliation{TU Dresden,
 Institute of Theoretical Physics and Center for Dynamics,
 01062 Dresden, Germany}

\date{\today}

\begin{abstract}
Resonance states spatially diverge and are thus not square integrable.
Instead, their norm is defined by the biorthogonal scalar product of left and
right states.
We replace the corresponding volume integral over infinite space by a
convenient boundary integral in piecewise-homogeneous systems and apply this
procedure to diverse physical settings.
For a quantum particle in any number of dimensions we treat hard-wall and
piecewise-constant potentials.
For electromagnetic systems with piecewise-homogeneous material properties, we
consider three-dimensional and effectively two-dimensional cavities.
\end{abstract}

\maketitle

\vspace*{1.cm}

\section{Introduction}

Scattering systems are important in various fields of physics and are studied
under the notion of non-Hermitian systems~\cite{Moi2011, AshGonUed2020}.
Fundamental to their description are resonance states (also called
resonant states, quasinormal modes, Gamow states, or leaky modes), which
are time-harmonic solutions to the wave equation of the system with
complex eigenfrequencies.
The negative imaginary part of the eigenfrequency describes the temporal decay of
a resonance state.
A (right) resonance state $\psiR$ is defined as a solution to the eigenvalue
equation $H | \psiR \rangle = \lambda \, | \psiR \rangle$,
with $H$ the operator describing the system and $\lambda$ the
complex eigenvalue.
For scattering systems in unbounded position space, the resonance states
fulfill an outgoing wave boundary condition $\ue^{\ui k |\vecr|}$ at infinity,
which implies due to the negative imaginary part of the wavenumber $k$
that they grow exponentially $\ue^{- \Imag(k) |\vecr|}$ for $|\vecr| \to \infty$.

For closed quantum systems, the norm of eigenstates is important for,
e.g., the computation of expectation values, the completeness relation,
and time evolution based on eigenstates.
Here, the suitable norm is the
$L^2$-norm, i.e., the square root of
$\langle \psi | \psi \rangle = \int_{\mathbb{R}^D} \dV \ |\psi(\vecr)|^2$.
For scattering systems, the normalization of resonance states is also
important for the above reasons~\cite{Moi2011, GolGilMoi2012}
and, e.g., the Purcell factor~\cite{SauHugMakLal2013, MulLan2016}, the Petermann
factor~\cite{Sch2009b, Wie2023, KulWie2025}, the phase
rigidity~\cite{LanBroBee1997, SadBer2005, RotBir2015, Wie2023, YiRyuRodHen2025},
the left-right Husimi representation~\cite{ErmCarSar2009, LorMoeKet2025},
and averaging over chaotic resonance states~\cite{HarShi2015, KulWie2016,
BitKimZenWanCao2020, KetClaFriBae2022, KetLorSch2025}.
However, the $L^2$-norm diverges due to the exponential growth of
resonance states for $|\vecr| \to \infty$.
Instead, the complex biorthogonal scalar product
$\langle \psiL | \psiR \rangle$ of left and
right resonance states is the relevant quantity~\cite{Zel1961, Moi2011,
SauWuZarMulLal2022}, which we call \textit{norm} for brevity, following
Refs.~\cite{Zel1961, SauWuZarMulLal2022}.
Note that it is not a norm in the mathematical sense.
Here, $\psiL$ is the left
resonance state fulfilling $\langle \psiL | H = \langle \psiL | \lambda$
with the same eigenvalue $\lambda$ as for the right resonance state
$\psiR$~\cite{Moi2011}.
Furthermore, one has biorthogonality of left and right resonance
states, i.e.,
$\langle \psiL_i | \psiR_j \rangle = 0$ for $i \neq j$~\cite{Ber1968,
Moi2011}.

\begin{figure}[b]
    \begin{center}
        \includegraphics{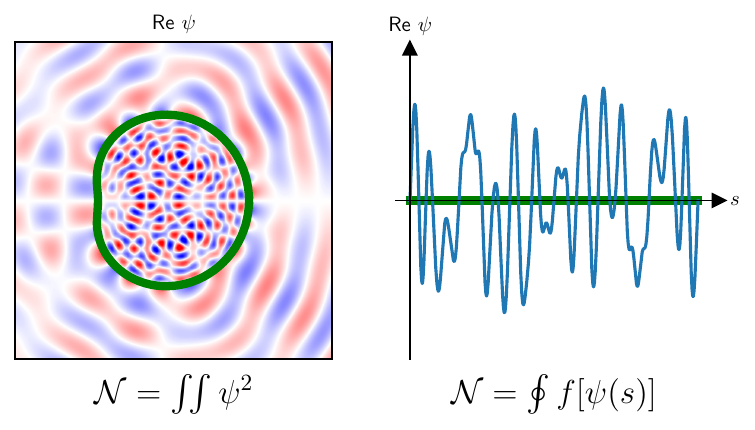}
    \end{center}
    \caption{
        Resonance state of a scattering system with piecewise-homogeneous
        properties showing its real part in position space (left) and on the boundary (right).
        While its norm is defined by an integral over infinite space, we show
        in various physical settings that it can be computed using a convenient boundary
        integral only.
        \vspace*{5cm}
    }
    \label{fig:overview}
\end{figure}

The above norm, $\langle \psiL | \psiR \rangle$,
of resonance states in scattering systems is given in
an unbounded position space by the integral
$\int_{\mathbb{R}^D} \dV \ \overline{\psiL(\vecr)} \, \psiR(\vecr)$.
The evaluation of this integral, however, faces two fundamental problems:
(i)~It turns out that the integrand containing left and right resonance
states also diverges for $|\vecr| \to \infty$ and
(ii)~the integral is over the whole space $\mathbb{R}^D$ and thus cumbersome
to evaluate for numerically determined resonance states.
Problem (i) can be resolved since the complex integrand oscillates and
its integral converges when applying regularization methods~\cite{Moi2011},
e.g., the Zel'dovich regularization~\cite{Zel1961, Ber1968,
StoColBonMcP2021}, complex scaling~\cite{Moi2011}, or analytical
continuation~\cite{Rom1968}.
A possible approach for problem (ii) is to replace the integral over the whole
space $\mathbb{R}^D$ by a volume integral over a finite region plus a boundary
integral over its surface.
This is achieved by writing the integrand as a divergence of a suitable vector
field and applying the divergence theorem, where the contribution from the
boundary at infinity vanishes.
Originally, this was done for one-dimensional electromagnetic
systems~\cite{LeuLiuYou1994} and for one-dimensional quantum scattering
systems~\cite{Lee2009}.
The first extension to more than one dimension was done for electromagnetic
systems~\cite{MulLanZim2010}.
The now established
definition of the norm for electromagnetic systems and its historical
development is reviewed in Ref.~\cite{SauWuZarMulLal2022}.
In these systems the norm can be expressed by an integral over a volume
including all inhomogeneities and a boundary integral over its
surface~\cite{MulLanZim2010,MulWei2018,StoColBonMcP2021,BotWei2022,SauWuZarMulLal2022}.
Among others this has been applied to two systems with analytically given
resonance states, the spherical scatterer (Mie scattering)~\cite{DooLanMul2014,
StoColBonMcP2021} and the effectively two-dimensional circular
disk~\cite{DooLanMul2013}.
Evaluating the norm of resonance states is also important in other areas
of physics, e.g., acoustic
systems~\cite{DomIgoSolSonLiPetBog2026:p, DomIgoSolSonLiPetBog2026:pb} and
black holes~\cite{JarMacShe2021}.

It is an open question how to address problem (ii) for quantum scattering
systems, in particular for more than one dimension.
For electromagnetic systems, it would be desirable to simplify the volume
integral for materials with homogeneous properties, e.g., for dielectric
cavities~\cite{CaoWie2015, JiaChaZhaWieHenWanCaoXiaAlu2026}.

In this paper, we derive convenient boundary integrals for the computation of the norm of resonance
states in quantum scattering systems and electromagnetic systems with piecewise
homogeneous properties bounded by piecewise-smooth boundaries,
visualized in \cref{fig:overview}.
For both kinds of systems the norm is based on the biorthogonality of left
and right resonance states and will be evaluated in position representation
using Zel'dovich regularization.
This gives a conceptually different derivation for the well-established norm in
electromagnetic systems.
Suitable vector fields allow for expressing the integrand of volume
integrals as a divergence, leading to boundary integrals for an efficient
evaluation of the norm.
Specifically, for a quantum particle in any number of dimensions we
give boundary integrals for the norm for hard-wall and piecewise-constant
potentials.
As examples, we apply them to the three-disk scattering system and
one-dimensional step potentials.
We derive a local volume integral for the norm for arbitrary potentials which
decay sufficiently fast.
For electromagnetic systems with piecewise-homogeneous material properties,
boundary integrals for the norm are derived in three dimensions and for
effectively two-dimensional cavities.
As analytical examples, we treat the spherical scatterer and the circular
disk.

The paper is structured as follows.
In \cref{sec:approach_results} we explain the general approach for the norm of
resonance states based on biorthogonality from non-Hermitian physics.
For quantum scattering systems we derive in \cref{sec:potential} expressions for
the norm in hard-wall, piecewise-constant, and arbitrary potentials.
For electromagnetic systems we rederive in \cref{sec:electromagnetic_systems}
the established norm and derive boundary integrals for piecewise-homogeneous
materials.
We summarize our main results for the norm expressed by boundary integrals in
\cref{sec:results}.

\FloatBarrier
\newpage

\section{General approach}%
\label{sec:approach_results}

In this section, we explain the general approach for evaluating the norm of
resonance states.
To this end, we first introduce the norm of resonance states based on
biorthogonality in non-Hermitian systems and discuss its definition in position space
using Zel'dovich regularization in \cref{sec:norm_biorthogonality}.
Then we describe the geometrical setup for the systems we consider in
\cref{sec:geometrical_setup}.
In \cref{sec:divergence} we define vector fields whose divergence appears in
volume integrals for the norm.
Finally, \cref{sec:divergence_theorem_zeldovich} treats the application of the
divergence theorem to integrals with Zel'dovich regularization.

\subsection{Norm and biorthogonality}%
\label{sec:norm_biorthogonality}

\subsubsection{Left and right resonance states}

Right and left resonance states of an operator $H$, e.g., describing a quantum particle in a scattering potential by the Schrödinger equation, see~\cref{sec:potential},
fulfill the eigenvalue equations
\begin{equation}
H | \psiR_i \rangle= \lambda_i \, | \psiR_i \rangle
\, , \qquad
\langle \psiL_i | H = \lambda_i \, \langle \psiL_i |
\, ,
\label{Eq:left_and_right_eigenvalue_equations}
\end{equation}
with outgoing wave boundary condition, see~\cref{Eq:asymptotic_form} below,
and incoming wave boundary condition, respectively.
As discussed in the introduction, left and right resonance
states are not square-integrable.
Thus they cannot be treated as elements of the
usual Hilbert space.
Instead, they are generalized eigenvectors in a rigged Hilbert space and are
only well-defined by their action on suitable test functions~\cite{DeGarMug2005,Mad2012}.
As a consequence, expressions like $\langle \psiL | \psiR \rangle$ have only
a formal meaning and do not imply the use of a Hilbert-space scalar
product~\cite{Mad2012}.
As first proposed by Zel'dovich~\cite{Zel1961}, such expressions may be
evaluated nevertheless by regularization, see
\cref{sec:zeldovich_regularization} below.
We here follow this convention and use the term biorthogonal scalar product for
brevity.
Note that the term \textit{left resonance state} is sometimes used with a
different meaning, see, e.g., Ref.~\cite{Mad2012}, where it is defined as
the Hermitian conjugate of the right resonance state.

Left and right resonance states are biorthogonal~\cite{Moi2011}
\begin{equation}
\langle \psiL_i | \psiR_j \rangle = \, \delta_{ij} \, \mathcal{N}_i
\, ,
\label{Eq:biorthogonality}
\end{equation}
i.e., all left resonance states are orthogonal to all right
resonance states, except if they belong to the same eigenvalue.
Then the biorthogonal scalar product $\langle \psiL_i | \psiR_i \rangle$ of
the $i$-th left and right resonance state
defines the complex value $\mathcal{N}_i \in \mathbb{C}$.
We call this the norm of the resonance state even though it is not a
norm in the mathematical sense, following the terminology in the
literature on normalization of resonance states~\cite{Zel1961,
SauWuZarMulLal2022}.
In this paper we assume for simplicity that there are no degeneracies like,
e.g., exceptional points.

For a nonlinear eigenvalue problem,
e.g., Maxwell equations for a material with frequency-dependent properties,
see~\cref{sec:electromagnetic_systems},
the eigenvalue equations can be written in the form
\begin{equation}
    T(\lambda_i) | \psiR_i \rangle = 0
    \, , \qquad
    \langle \psiL_i | T(\lambda_i) = 0
    \, ,
    \label{Eq:nonlinear_left_and_right_eigenvalue_equations}
\end{equation}
where the operator $T$ depends nonlinearly on the eigenvalue $\lambda_i$
in contrast to the linear case $T(\lambda_i)= \lambda_i \mathbb{1} - H$ corresponding to \cref{Eq:left_and_right_eigenvalue_equations}.
Here, we conjecture that left and right resonance states are biorthogonal
with respect to the generalized biorthogonal scalar product
\begin{equation}
     \langle \psiL_i | D_{ij} | \psiR_j \rangle = \, \delta_{ij} \, \mathcal{N}_i
    \quad \text{with} \quad
    D_{ij} =
    \begin{cases}
        T'(\lambda_i)
        \, , & i = j
        \vspace{0.2cm} \\
        \dfrac{T(\lambda_i) - T(\lambda_j)}{\lambda_i - \lambda_j}
        \, , &  i \neq j
    \end{cases}
    \quad .
    \label{Eq:nonlinear_biorthogonality}
\end{equation}
The difference quotient for $i \neq j$ follows directly from the eigenvalue equations
and makes the appearance of the derivative $T'(\lambda_i)$ for $i=j$ plausible.
For the linear case $T(\lambda_i)= \lambda_i \mathbb{1} - H$ the operator
$D_{ij} = \mathbb{1}$ reduces to the identity.
The above generalized scalar product was derived for Hermitian operators
on a Hilbert space~\cite[Eqs.~(20) and (21)]{Had1968}, \cite[Sec.~60.3]{Vos2013}.
Our conjecture extends this to non-Hermitian operators with left and right
resonance states.
As discussed after \cref{Eq:left_and_right_eigenvalue_equations}, one has to evaluate the
left-hand side of the above equation by
regularization, see \cref{sec:zeldovich_regularization} below.
In \cref{sec:biorthogonality_electromagnetic_systems}, we demonstrate that
biorthogonality with respect to this generalized biorthogonal scalar product is indeed
fulfilled for the example of electromagnetic systems.

The norm $\mathcal{N}_i$ of the $i$-th resonance state can be used for any desired normalization,
typically by scaling right and left resonance states,
\begin{equation}
    | \hat{\psi}^{\: \mathrm{R}}_i \rangle = \frac{1}{\sqrt{\mathcal{N}_i}} | \psiR_i \rangle
    \quad \text{and} \quad
    \langle \hat{\psi}^{\: \mathrm{L}}_i | = \frac{1}{\sqrt{\mathcal{N}_i}} \langle \psiL_i |
    \, ,
\end{equation}
such that after scaling their norm equals one,
$\hat{\mathcal{N}}_i = \langle \hat{\psi}^{\: \mathrm{L}}_i
| \hat{\psi}^{\: \mathrm{R}}_i \rangle = 1$.

\subsubsection{Position representation}

For a resonance state the outgoing wave boundary condition at infinity is given
in position representation by
\begin{equation}
    \psiR_i(\vecr) = \langle \vecr | \psiR_i \rangle \sim h_i(\boldsymbol{\varphi}, k_i) \ r^{-\frac{D-1}{2}} \ \ue^{\ui k_i r}
    \quad
    \text{for}
    \quad
    r \to \infty \, ,
    \label{Eq:asymptotic_form}
\end{equation}
where $r = |\vecr|$ and $h_i(\boldsymbol{\varphi}, k)$ is the angular
distribution in the farfield of a $D$-dimensional space.
The wavenumber $k_i$ is complex with $\Imag\,k_i < 0$ and its relation to the
eigenvalue $\lambda_i$ depends on the operator.
As a consequence, the absolute squared right resonance state grows exponentially like
$|\psiR_i(\vecr)|^2 \sim r^{-(D-1)} \ue^{- 2 \Imag(k_i) r}$ for $r \to \infty$, such that
the $L^2$-norm diverges.

The eigenvalue equations, \cref{Eq:left_and_right_eigenvalue_equations},
can also be written in position representation
and we assume that our operators of interest are local in position representation,
such that we have
\begin{equation}
    H(\vecr) \, \psiR_i(\vecr) = \lambda_i \, \psiR_i(\vecr)
    \, , \qquad
    H^\top\!(\vecr) \, \overline{\psiL_i(\vecr)} = \lambda_i \, \overline{\psiL_i(\vecr)}
    \, ,
    \label{Eq:eigenvalue_equations_position}
\end{equation}
where in the second equation the transposed operator $H^\top$ in position representation
acts to the right and the complex conjugated left resonance state is defined by
$\overline{\psiL_i(\vecr)} = \langle \psiL_i | \vecr \rangle$.
For a nonlinear eigenvalue problem, \cref{Eq:nonlinear_left_and_right_eigenvalue_equations},
the eigenvalue equations in position representation are
\begin{equation}
    T(\lambda_i, \vecr) \, \psiR_i(\vecr) =  0
    \, , \qquad
    T^\top\!(\lambda_i, \vecr) \, \overline{\psiL_i(\vecr)} = 0
    \, .
    \label{Eq:nonlinear_eigenvalue_equations_position}
\end{equation}

Left resonance states $\psiL_i(\vecr)$ obey incoming wave boundary
conditions at infinity, such that
complex conjugated left resonance states
$\overline{\psiL_i(\vecr)}$
obey the same outgoing wave boundary conditions
as right resonance states $\psiR_i(\vecr)$, \cref{Eq:asymptotic_form}.
If the given operator is symmetric, $H^\top = H$ or $T^\top = T$,
one finds that $\overline{\psiL_i(\vecr)}$ and $\psiR_i(\vecr)$ fulfill the same
eigenvalue equation, making them proportional to each other and allowing the
natural choice
\begin{equation}
    \overline{\psiL_i(\vecr)} = \psiR_i(\vecr)
    \, ,
    \label{Eq:left_right_relation}
\end{equation}
i.e., in position representation left resonance states are the complex conjugate
of right resonance states~\cite{SteWal1972, Moi2011}.

\subsubsection{Zel'dovich regularization}%
\label{sec:zeldovich_regularization}

Evaluating the biorthogonal scalar product,
\cref{Eq:biorthogonality,Eq:nonlinear_biorthogonality},
in position representation leads to integrals where the
integrand has an exponentially increasing modulus
$\sim |\ue^{2 \ui k_i r}| = \ue^{-2 \Imag(k_i) r}$ and cannot be evaluated directly.
As the phase of the integrand oscillates,
the integral might still converge.
It is well-defined by suitable regularization methods~\cite{Moi2011}, e.g., the
Zel'dovich regularization~\cite{Zel1961, Ber1968, StoColBonMcP2021}, complex
scaling~\cite{Moi2011}, or analytical continuation~\cite{Rom1968}.

Here, we use the Zel'dovich regularization
for the biorthogonal scalar product,
\cref{Eq:biorthogonality}, giving
\begin{align}
    \langle \psiL_i | \psiR_j \rangle
    &=
    \lim_{\eta \to 0^{+}}
    \int_{\mathbb{R}^D} \dV \
    \overline{\psiL_i(\vecr)} \ \psiR_j(\vecr) \
    \ue^{-\eta r^2} \nonumber \\
    &=
    \lim_{\eta \to 0^{+}}
    \int_{\mathbb{R}^D} \dV \
    \psiR_i(\vecr) \ \psiR_j(\vecr) \
    \ue^{-\eta r^2}
    \quad (H = H^\top)
    \, ,
    \label{Eq:biorthogonality_regularized}
\end{align}
where the Gaussian $\ue^{-\eta r^2}$ ensures that the integrand
vanishes at infinity.
In the second line we consider \cref{Eq:left_right_relation} for a symmetric operator $H$.
Note that the limit and the integral cannot be interchanged.
For the nonlinear case the biorthogonal scalar product,
\cref{Eq:nonlinear_biorthogonality}, is given by
\begin{align}
    \langle \psiL_i | D_{ij} |\psiR_j \rangle
    &=
    \lim_{\eta \to 0^{+}}
    \int_{\mathbb{R}^D} \dV \
    \overline{\psiL_i(\vecr)} \, D_{ij} \, \psiR_j(\vecr) \
    \ue^{-\eta r^2} \nonumber \\
    &=
    \lim_{\eta \to 0^{+}}
    \int_{\mathbb{R}^D} \dV \
    \psiR_i(\vecr) \, D_{ij} \, \psiR_j(\vecr) \
    \ue^{-\eta r^2}
    \quad (T = T^\top)
    \, .
    \label{Eq:nonlinear_biorthogonality_regularized}
\end{align}
In particular, the diagonal terms $i=j$ give the norm,
\begin{align}
    \mathcal{N}_i
    & =
    \lim_{\eta \to 0^{+}}
    \int_{\mathbb{R}^D} \dV \
    \psiR_i(\vecr) \, T'(\lambda_i) \, \psiR_i(\vecr) \
    \ue^{-\eta r^2} \, ,
    \label{Eq:nonlinear_norm_regularized}
\end{align}
which for the linear eigenvalue problem reduces to
\begin{align}
    \mathcal{N}_i
    & =
    \lim_{\eta \to 0^{+}}
    \int_{\mathbb{R}^D} \dV \
    \left[\psiR_i(\vecr)\right]^2 \
    \ue^{-\eta r^2} \, .
    \label{Eq:norm_regularized}
\end{align}
Since we consider symmetric operators only, we need for
\cref{Eq:biorthogonality_regularized,Eq:nonlinear_biorthogonality_regularized,Eq:norm_regularized,Eq:nonlinear_norm_regularized}
only the right resonance states~$\psiR_i(\vecr)$.
Therefore, we omit the superscript `R' in most of the following with
$\psi_i(\vecr)$ denoting a right resonance state.

\subsection{Geometrical setup}%
\label{sec:geometrical_setup}

We consider in an infinite $D$-dimensional space a finite region $\Gamma$,
which can be a union
\begin{equation}
    \Gamma = \bigcup_{l=1}^n \Gamma_l
    \label{Eq:region_gamma}
    \, ,
\end{equation}
of $n$ unconnected regions.
The region $\Gamma$ has a piecewise-smooth $(D-1)$-dimensional boundary $\partial \Gamma$,
which at position $\vecr$ has an infinitesimal surface element $\dS$
with unit normal vector $\nuVec(\vecr)$ pointing outwards of $\Gamma$.
Along the boundary the gradient $\nabla$ can be decomposed into a normal and a tangential component,
\begin{equation}
    \nabla = \nuVec(\vecr) \, \partial_\nu + \nabla_{\tVec}
    \label{Eq:nabla_normal_tangential}
    \, ,
\end{equation}
with the normal derivative
\begin{align}
    \partial_{\nu}
    &= \nuVec(\vecr) \cdot \nabla
    \label{Eq:normal_derivative}
    \, .
\end{align}

\subsection{Suitable vector fields and their divergences}%
\label{sec:divergence}

In order to simplify $D$-dimensional integrals into
$(D-1)$-dimensional boundary integrals
we write the integrand as the divergence of a suitable vector field.
This is done for the three different integrands
$\psi^2(\vecr)$,
$\psi_i(\vecr) \, \psi_j(\vecr)$ with $i \ne j$,
and $|\psi(\vecr)|^2$.
Further cases specific to electromagnetic systems are given in
\cref{sec:electromagnetic_systems}.

\subsubsection{Integrand $\psi^2(\vecr)$}

We define a vector field in $D$ dimensions
\begin{align}
    \RVec(\vecr) &= \frac{1}{2} \, \vecr \, \psi\psi
    + \frac{1}{2 \lambda} \,
    \Big[
    \left(D - 2 \right) \psi \, \nabla \psi
    - \vecr \, \left( \nabla \psi \cdot \nabla \psi \right)
    + 2 (\vecr \cdot \nabla \psi) \, \nabla \psi
    \Big]
    \, ,
    \label{Eq:R_definition}
\end{align}
with $\lambda \in \mathbb{C}$.
The divergence of $\RVec(\vecr)$
depends on the eigenvalue equation satisfied by $\psi$,
\begin{align}
    -\Delta\psi & = \lambda \psi
    \quad \Rightarrow \quad
    \nabla \cdot \RVec
    = \psi^2
    \label{Eq:R_divergence}
    \\
    \left[-\Delta + V(\vecr) \right] \psi & = \lambda \psi
    \quad \Rightarrow \quad
    \nabla \cdot \RVec
    = \psi^2
    + \frac{V \psi}{2 \lambda} \Big[\left(D - 2 \right) \psi + 2 \vecr \cdot \nabla \psi \Big]
    \label{Eq:R_divergence_potential}
    \, ,
\end{align}
see \cref{sec:appendix_divergence} for the derivation.
Note that the vector field $\RVec(\vecr)$ depends on the choice of the
origin, while its divergence, \cref{Eq:R_divergence}, does not.

The vector field $\RVec$ was implicitly used by Rellich~\cite[Eq.~(2)]{Rel1940}
to derive a boundary integral for the eigenvalue $\lambda$ in $(\Delta + \lambda) \psi = 0$
for Dirichlet boundary conditions.
For $D = 3$ it was independently introduced for the
normalization of resonance states in electromagnetic systems by Muljarov and
Langbein~\cite[Eq.~(B27)]{MulLan2016}.

\subsubsection{Integrand $\psi_i(\vecr) \, \psi_j(\vecr)$ with $i \ne j$}

We define a vector field in $D$ dimensions
for different functions $\psi_i(\vecr)$ and $\psi_j(\vecr)$ with $i \ne j$,
    \begin{equation}
        \BVec_{ij}(\vecr)
        =
        \frac{\psi_i \, \nabla \psi_j - \psi_j \, \nabla \psi_i}{\lambda_i - \lambda_j}
        \, ,
        \label{Eq:L_definition}
    \end{equation}
with some constants $\lambda_i , \lambda_j \in \mathbb{C}$ and
$\lambda_i \neq \lambda_j$.
This vector field was previously used to show the biorthogonality of resonance
states in one-dimensional systems by Berggren~\cite{Ber1968} % Eq. after Eq. (2.11) in Ber1968
and of resonance states in electromagnetic systems by Muljarov and Langbein~\cite[Eq.~(B28)]{MulLan2016}.
We obtain the divergence of $\BVec_{ij}(\vecr)$
    \begin{align}
        \nabla \cdot \BVec_{ij}
        &=
        \frac{\psi_i \, \Delta \psi_j - \psi_j \, \Delta \psi_i}{\lambda_i - \lambda_j}
        =
        \psi_i \, \psi_j
        \, ,
        \label{Eq:L_divergence}
    \end{align}
where in the last step we use that $\psi_i$ and $\psi_j$ both satisfy
either the eigenvalue equation
$- \Delta\psi_{i/j} = \lambda_{i/j} \psi_{i/j}$
or
$\left[-\Delta + V(\vecr) \right] \psi_{i/j} = \lambda_{i/j} \psi_{i/j}$.

\subsubsection{Integrand $|\psi(\vecr)|^2$}%
\label{sec:divergence_abs_psi}

We mention for completeness the case of an integrand
$|\psi|^2 = \overline{\psi} \psi$,
which differs from $\psi^2$ if the scalar field $\psi$ is complex.
Here, the choice of the vector field depends on whether the eigenvalue $\lambda$ is
complex or real.
\begin{itemize}
    \item[(i)]  For complex $\lambda \in \mathbb{C}$ with
        $\text{Im} \, \lambda \ne 0$, as it is the case for resonance states,
        one can exploit the continuity equation and define the vector field
        \begin{align}
            \JVec(\vecr)
            &=
            \frac{\text{Im} \left[ \, \overline{\psi}(\vecr) \ \nabla\psi(\vecr) \right]}
                 {- \text{Im} \, \lambda}
            \, ,
        \end{align}
        where the numerator is proportional to a probability current and
        for $\lambda = k^2$
        the denominator $\text{Im} \, k^2 = 2 \, \text{Im}\,k \, \text{Re}\,k$
        is proportional to the decay rate.
        The divergence of $\JVec(\vecr)$ is
        \begin{align}
            \nabla \cdot \JVec
            &=
            \frac{\text{Im} \left[ \nabla \overline{\psi} \cdot \nabla\psi + \overline{\psi} \ \Delta\psi \right]}
                 {- \text{Im} \, \lambda}
            =
            \frac{\text{Im} \left[ \, \overline{\psi} \ \Delta\psi \right]}{- \text{Im} \, \lambda}
            = |\psi|^2
            \, ,
        \end{align}
        where in the last step we use that $\psi$ satisfies
        either the eigenvalue equation
        $-\Delta\psi = \lambda \psi$
        or
        $\left[-\Delta + V(\vecr) \right] \psi = \lambda \psi$
        with a real potential $V(\vecr)$.

    \item[(ii)] For real $\lambda \in \mathbb{R}$
        we define a vector field in $D$ dimensions
        \begin{align}
            \RVecTilde(\vecr) &= \frac{1}{2} \, \vecr \, \overline{\psi} \psi
            + \frac{1}{4 \lambda} \,
            \Big[
            \left(D - 2 \right) \overline{\psi} \, \nabla \psi
            - \vecr \, \left( \nabla \overline{\psi} \cdot \nabla \psi \right)
            + 2 (\vecr \cdot \nabla \overline{\psi}) \, \nabla \psi
            + \text{comp.\ conj.}
            \Big]
            \label{Eq:R_tilde_definition}
            \, ,
        \end{align}
        where $\RVecTilde(\vecr)$ is inspired by \cref{Eq:R_definition}.
        A similar vector field was derived by
        Boasman~\cite[Eq.~(154)]{Boa1994} for closed billiards in $D=2$
        dimensions and $\lambda = k^2 \in \mathbb{R}$, which however contains
        second derivatives of $\psi$.
        This makes $\RVecTilde$ preferable for most applications.
        The divergence of $\RVecTilde(\vecr)$
        for solutions $\psi$ to the eigenvalue equation
        $-\Delta\psi = \lambda \psi$ with $\lambda \in \mathbb{R}$ is
        \begin{align}
            \nabla \cdot \RVecTilde
            = |\psi|^2
            \, ,
        \end{align}
        see \cref{sec:appendix_divergence}.
\end{itemize}

\subsection{Divergence theorem and Zel'dovich regularization}%
\label{sec:divergence_theorem_zeldovich}

In this section we show how to apply the divergence theorem to integrals with
Zel'dovich regularization, which often appear in this paper.
We find that a $D$-dimensional volume integral outside a region $\Gamma$ can be
evaluated by a $(D-1)$-dimensional boundary integral on $\partial \Gamma$,
\begin{equation}
    \lim_{\eta \to 0^{+}}
    \int_{\mathbb{R}^D \setminus \Gamma} \dV \
    \big[ \nabla \cdot \boldsymbol{f}(\vecr) \big] \
    \ue^{-\eta r^2}
    =
    - \oint_{\partial \Gamma} \dS \ \nuVec(\vecr) \cdot
    \boldsymbol{f}^\mathrm{\, out}(\vecr) \, .
    \label{Eq:regularized_integral_divergence}
\end{equation}
Examples for such a vector field $\boldsymbol{f}(\vecr)$ are $\RVec(\vecr)$ and
$\BVec(\vecr)$ from \cref{sec:divergence}.
If $\boldsymbol{f}(\vecr)$ is discontinuous across the boundary $\partial
\Gamma$, the value of $\boldsymbol{f}(\vecr)$ at the boundary is obtained
from the limit when approaching $\partial \Gamma$ from the outside of $\Gamma$
and is thus denoted by $\boldsymbol{f}^\mathrm{\, out}(\vecr)$.
If there is no region $\Gamma$ the integral is taken over the
entire space $\mathbb{R}^D$ and the right-hand side is absent,
\begin{equation}
    \lim_{\eta \to 0^{+}}
    \int_{\mathbb{R}^D} \dV \
    \big[ \nabla \cdot \boldsymbol{f}(\vecr) \big] \
    \ue^{-\eta r^2}
    =
    0
    \, .
    \label{Eq:regularized_integral_divergence_no_Gamma}
\end{equation}
The same result can be derived if there is a boundary
$\partial \Gamma$ within $\mathbb{R}^D$ where $\nuVec \cdot \boldsymbol{f}$ is
continuous across the boundary, such that boundary contributions from both
sides of $\partial \Gamma$ cancel.

The contribution from the boundary at infinity vanishes in
\cref{Eq:regularized_integral_divergence,Eq:regularized_integral_divergence_no_Gamma}
under the following assumptions:
\begin{itemize}
\item[(i)] The vector field $\boldsymbol{f}(\vecr)$ depends on products of resonance states of the form
$\psi_i(\vecr) \, \psi_j(\vecr)$ or products of their spatial derivatives.

\item[(ii)] The resonance states $\psi_i(\vecr)$ and $\psi_j(\vecr)$ have the
asymptotic form given by \cref{Eq:asymptotic_form} with
$\Real \, k_{i/j} > 0$ and $\Imag \, k_{i/j} < 0$.

\item[(iii)] The resonance poles $k_i$ and $k_j$ fulfill
$\Real \, k_{i/j} > - \Imag \, k_{i/j}$, see discussion at the end of
this section.

\item[(iv)] The vector field $\boldsymbol{f}(\vecr)$ may have an explicit polynomial
dependence on the position $\vecr$.
\end{itemize}

We derive \cref{Eq:regularized_integral_divergence} by writing the integrand as
\begin{equation}
    \big[ \nabla \cdot \boldsymbol{f}(\vecr) \big] \
    \ue^{-\eta r^2}
    =
    \nabla \cdot \left[ \boldsymbol{f}(\vecr) \, \ue^{-\eta r^2} \right]
    + 2\eta \, \vecr \cdot \boldsymbol{f}(\vecr) \, \ue^{-\eta r^2}
    \, ,
\end{equation}
and by applying the divergence theorem to the first term, giving
\begin{align}
    \lim_{\eta \to 0^{+}} \int_{\mathbb{R}^D \setminus \Gamma} \dV \
    \big[ \nabla \cdot \boldsymbol{f}(\vecr) \big] \
    \ue^{-\eta r^2}
    = &
    \lim_{\eta \to 0^{+}}
    \oint_{\partial \Gamma} \dS \ [-\nuVec(\vecr)] \cdot \boldsymbol{f}(\vecr) \,
    \ue^{-\eta r^2}
    \label{Eq:integral_domain_decomposition}
    \nonumber
    \\
    &
    +
    \lim_{\eta \to 0^{+}} \,
    \underbrace{\lim_{R \to \infty} \oint_{\partial B_R} \dS \
    \nuVec(\vecr) \cdot \boldsymbol{f}(\vecr) \, \ue^{-\eta R^2}}_{=0}
    \nonumber \\
    &
    \ + \
    \underbrace{
    2 \lim_{\eta \to 0^{+}} \,
    \eta \int_{\mathbb{R}^D \setminus \Gamma} \dV \ \vecr \cdot
    \boldsymbol{f}(\vecr) \,
    \ue^{-\eta r^2}}_{=0} \, .
\end{align}
Here in the first term $\nuVec(\vecr)$ points outward of the region
$\Gamma$, see \cref{sec:geometrical_setup}, so that $-\nuVec(\vecr)$
points outward of the integration region $\mathbb{R}^D \setminus \Gamma$, as
required by the divergence theorem.
Taking the limit $\eta \to 0^{+}$ yields the boundary
integral on the right-hand side in \cref{Eq:regularized_integral_divergence},
where $\boldsymbol{f}(\vecr)$ is evaluated on the outside of $\partial \Gamma$.

The second term is the boundary integral at infinity.
It is given by a $(D-1)$-dimensional surface $\partial B_R$ of a $D$-dimensional ball $B_R$ with radius $R$ in the limit $R \to \infty$.
It vanishes for any $\eta > 0$ in this limit due to the faster decay of the
Gaussian factor compared to the asymptotic growth of the resonance states
$\psi_i(\vecr)$ and $\psi_j(\vecr)$,~\cref{Eq:asymptotic_form}.

The third term vanishes for $\eta \to 0^{+}$ for the following reason:
The vector field $\boldsymbol{f}(\vecr)$ depends on the resonance states
$\psi_i(\vecr)$ and $\psi_j(\vecr)$, which grow exponentially for $r \to \infty$.
For small $\eta$ the dominant contribution to the integral comes
from large $r$ where the asymptotic
form, \cref{Eq:asymptotic_form}, is valid.
As a consequence, the angular integration over $D-1$ dimensions can be
performed separately
and one is left with a radial integral that has the asymptotic form
\begin{equation}
    \lim_{\eta \to 0^{+}} \eta \int_a^\infty \ud r \, r^s \,
    \ue^{\ui \tilde{k} r} \ue^{-\eta r^2},
    \label{Eq:radial_integral}
\end{equation}
where $a > 0$ is a lower bound.
The value of the exponent $s \in \mathbb{R}$ depends on the specific form of
the vector field $\boldsymbol{f}(\vecr)$.
Here one has $\tilde{k} = 2 k_i$ for evaluating the norm
and $\tilde{k} = k_i + k_j$ for confirming biorthogonality.
In both cases, it follows from assumption (ii) that $\Real \, \tilde{k} > 0$
and $\Imag \, \tilde{k} < 0$, and from assumption (iii) that $\Real \, \tilde{k} > |\Imag \, \tilde{k}|$.
As shown in \cref{sec:appendix_radial_integral},
the radial integral \cref{Eq:radial_integral} vanishes for $\eta \to 0^{+}$.
This argument for the leading asymptotic term can be repeated for all
subleading terms of the asymptotic expansion, which have a smaller
exponent $s$.

Assumption (iii) is fulfilled by many resonance states, e.g., by long-lived
resonances with small $|\Imag \, k|$ or in the semiclassical limit
$\Real \, k \to \infty$.
This assumption appears as a consequence of using Zel'dovich regularization
with the Gaussian factor $\ue^{-\eta r^2}$ and does not occur in other
regularization methods~\cite{Moi2011}.
Alternatively, using a generalized Zel'dovich regularization with a different
regularization function $\ue^{-\eta r^p}$ with suitable $p > 1$ allows for
dropping assumption (iii)~\cite{LorPhD}.
Therefore, all results in this work apply to resonances satisfying $\Real \, k > 0$ and $\Imag \, k < 0$ according to assumption (ii).

\newpage

\section{Potential scattering}%
\label{sec:potential}

In this section, we consider a quantum particle in a $D$-dimensional potential
$V(\vecr)$ described by the Hamiltonian
\begin{equation}
    H = - \Delta + V(\vecr)\, ,
    \label{Eq:hamiltonian_potential}
\end{equation}
where we set $\frac{\hbar^2}{2m} = 1$.
In position representation, right and left resonance states are defined by
\cref{Eq:eigenvalue_equations_position}.
They are related by complex conjugation, \cref{Eq:left_right_relation}, as the
Hamiltonian $H$ is symmetric.
Thus we need the eigenvalue equation for
right resonance states $\psi_i(\vecr)$ only,
which is given by
\begin{equation}
    \left[-\Delta + V(\vecr) \right] \psi_i(\vecr) = k_i^2 \psi_i(\vecr)
    \label{Eq:helmholtz_right_potential}
    \, ,
\end{equation}
where the eigenvalue $\lambda_i = k_i^2$ is expressed by the wavenumber $k_i$
appearing in the outgoing boundary condition, \cref{Eq:asymptotic_form}.

First, we give a short verification of biorthogonality under Zel'dovich regularization, i.e.,
$\langle \psiL_i | \psiR_j \rangle = 0$ for $i \neq j$.
To this end, we use that in~\cref{Eq:biorthogonality_regularized} for $i \neq j$ the
integrand
$\psiR_i(\vecr) \psiR_j(\vecr) = \nabla \cdot \BVec_{ij}(\vecr)$
can be written as a divergence of the vector field $\BVec_{ij}(\vecr)$,
\cref{Eq:L_definition} with $\lambda_{i/j} = k_{i/j}^2$ and
$\psi_{i/j} = \psiR_{i/j}$.
We apply the divergence theorem to \cref{Eq:biorthogonality_regularized}.
For hard-wall potentials we use
\cref{Eq:regularized_integral_divergence}
with $\boldsymbol{f}(\vecr)=\BVec_{ij}(\vecr)$, which vanishes on the
boundary $\partial \Gamma$.
For piecewise-constant potentials and for arbitrary smooth potentials we use
\cref{Eq:regularized_integral_divergence_no_Gamma} with
$\boldsymbol{f}(\vecr)=\BVec_{ij}(\vecr)$.
In all cases the integral vanishes, giving biorthogonality.

In the following, we show that the norm of a resonance state can be expressed via a
$(D-1)$-dimensional boundary integral for hard-wall and piecewise-constant
potentials, see \cref{sec:hard_wall_potential,sec:piecewise_constant_potential},
while for arbitrary, sufficiently fast decaying potentials it can be computed by a
local $D$-dimensional integral, see \cref{sec:smooth_potential}.

\subsection{Hard-wall potential}%
\label{sec:hard_wall_potential}

In the case of a hard-wall potential,
\begin{equation}
    V(\vecr) =
    \begin{cases}
        \infty \, , & \vecr \in \Gamma \\
        0 \, , & \text{otherwise}
    \end{cases}
    \quad ,
\end{equation}
resonance states $\psi(\vecr)$ fulfill Dirichlet boundary conditions at the
boundary $\partial \Gamma$,
\begin{equation}
    \psi(\vecr)
    = 0 \quad \text{for} \quad
    \vecr \in \partial \Gamma \;
    \label{Eq:boundary_conditions_wf_obstacle}
\end{equation}
in addition to the outgoing wave boundary conditions, \cref{Eq:asymptotic_form},
at infinity.
As a consequence, the gradient $\nabla \psi$ has no tangential component
along the boundary, but is given just by its normal component
\begin{equation}
    \nabla \psi
    =
    \partial_{\nu} \psi \, \nuVec(\vecr) \; ,
    \label{Eq:gradient_obstacle}
\end{equation}
with the normal derivative $\partial_{\nu} \psi$ defined in \cref{Eq:normal_derivative}.

\subsubsection{Norm}%
\label{sec:obstacle_diagonal}

We show that the norm $\mathcal{N}_i$ of a resonance state
can be computed by a boundary integral.
To this end, we use that in~\cref{Eq:norm_regularized} the integrand
$\left[\psi_i(\vecr) \right]^2 = \nabla \cdot \RVec(\vecr)$ can be written as
a divergence of the vector field $\RVec(\vecr)$, \cref{Eq:R_definition}
with $\lambda = k_i^2$ and $\psi = \psi_i$.
Applying the divergence theorem to~\cref{Eq:norm_regularized}, according
to~\cref{Eq:regularized_integral_divergence} with
$\boldsymbol{f}(\vecr)=\RVec(\vecr)$, we find for the norm
\begin{equation}
    \mathcal{N}_i =
    - \oint_{\partial \Gamma} \dS \
    \nuVec(\vecr) \cdot \RVec^\mathrm{\, out}(\vecr) \, .
    \label{Eq:integral_domain_obstacle_norm}
\end{equation}
Using \cref{Eq:R_definition} and Dirichlet boundary conditions,
\cref{Eq:boundary_conditions_wf_obstacle,Eq:gradient_obstacle}, we find for the
norm
\begin{align}
    \mathcal{N}_i
    &=
    -\frac{1}{2 k_i^2} \,
    \oint_{\partial \Gamma} \dS \
    \left[ \nuVec(\vecr) \cdot \vecr \right] \,
    \left[ \partial_{\nu} \psi_i(\vecr) \right]^2
    \label{Eq:norm_hard_wall}
    \, .
\end{align}
Interestingly, this resembles the expression for the $L^2$-norm of an eigenstate
of a closed billiard with Dirichlet boundary conditions up to a minus
sign~\cite{Rel1940, BerWil1984}.

Note that this boundary integral does not depend on the choice of the origin of
the coordinate system, even though the integrand depends on the origin.
This is due to the fact that the divergence of $\RVec(\vecr)$,
\cref{Eq:R_divergence}, does not depend on the choice of the origin.
Thus, the origin can be chosen such that its evaluation becomes
as simple as possible, e.g., such that $\nuVec(\vecr) \cdot \vecr$ vanishes or
is constant on some part of the boundary.

\subsubsection{Example: Three-disk scattering system}%
\label{sec:three_disk_scattering}

We illustrate the evaluation of the boundary integral for the
norm, \cref{Eq:norm_hard_wall}, for
the three-disk scattering system.
This is a paradigmatic model in quantum chaotic scattering and has been
studied extensively~\cite{CviEck1989, GasRic1989c, Smi1989,
Wir1999, CviArtMaiTan2020, WeiBarKuhPolSch2014, SchKet2023, KetLorSch2025}.
In this system, resonance states are numerically determined by a Fourier series
expansion on the disks' boundaries.
Therefore, it is desirable to express the norm in terms of the Fourier coefficients.

The two-dimensional ($D=2$) three-disk scattering system consists of three
disks $l=1,2,3$ of radius $a$ on vertices of an equilateral triangle of side
length $R$ at positions $\boldsymbol{c}_l$, which make up the region $\Gamma$.
The boundary $\partial \Gamma_l$ of each disk $l$ is described
by $\vecr_l = \boldsymbol{c}_l + a \nuVec_l(\theta_l)$
using the local polar angle $\theta_l$~\cite[Fig.~2]{GasRic1989c}
with $\nuVec_l(\theta_l)$ the outward pointing normal vector.
Thus, we find for the factor $\nuVec \cdot \vecr$ appearing
in~\cref{Eq:norm_hard_wall},
\begin{equation}
    \nuVec_l \cdot \vecr_l = \nuVec_l \cdot \boldsymbol{c}_l + a
    = a + \frac{R}{\sqrt{3}} \cos \theta_l,
\end{equation}
where we used that $| \boldsymbol{c}_l | = R / \sqrt{3}$.

For the evaluation of~\cref{Eq:norm_hard_wall},
we use that the normal derivative on the boundary of disk $l$ is expressed in
terms of a Fourier series expansion by~\cite{GasRic1989c}
\begin{equation}
    \partial_{\nu} \psi_i(\vecr_l) = \frac{1}{a^2}
    \sum_{m=-\infty}^{\infty} A_{ilm} \ue^{\ui m \theta_l} \, ,
\end{equation}
with the $m$-th Fourier coefficients $A_{ilm}$ of the $i$-th resonance state at
disk $l$.
Using further that $\dS = a \,  \ud \theta_l$ on the boundary of disk
$l$, we find for the norm,
\cref{Eq:norm_hard_wall},
\begin{align}
    \mathcal{N}_i
    =& - \frac{1}{2 k_i^2 a^2}
    \sum_{l=1}^3
    \sum_{m,n=-\infty}^{\infty} A_{ilm} A_{iln}
    \int_0^{2\pi} \ud \theta_l \left( 1 + \frac{R}{\sqrt{3} a}
    \cos \theta_l \right) \ue^{\ui (m+n) \theta_l}
    \\
    =& - \frac{\pi}{k_i^2 a^2}
    \sum_{l=1}^3
    \sum_{m=-\infty}^{\infty} A_{ilm}
    \bigg[ A_{il,-m} + \frac{R}{2 \sqrt{3} a}
    \left( A_{il,-m-1} + A_{il,-m+1} \right) \bigg] \,.
\end{align}

By making use of the symmetries of the three-disk system, one can
classify the resonance states according to the irreducible
representations of the $C_{3\mathrm{v}}$ symmetry
group~\cite{GasRic1989c}, which leads to further simplifications of the
norm:
\begin{enumerate}
    \item[(i)] $A_1$ representation:
        $A_{i1m} = A_{i2m} = A_{i3m} \equiv A_{im}$ and $A_{im} = A_{i,-m}$
        \begin{align}
            \implies
            \mathcal{N}_i^{A_1}
            &= \frac{6 \pi}{k_i^2 a^2}
            \left(
            \frac{1}{2} A_{i0}^2
            - \sum_{m=0}^{\infty}
            \bigg[
             A_{im}^2
            + \frac{R}{\sqrt{3} a} A_{im} A_{i,m+1}
            \bigg]
            \right) \, .
        \end{align}

    \item[(ii)] $A_2$ representation:
        $A_{i1m} = A_{i2m} = A_{i3m} \equiv A_{im}$ and $A_{im} = -A_{i,-m}$
        \begin{align}
            \implies
            \mathcal{N}_i^{A_2}
            &= \frac{6 \pi}{k_i^2 a^2}
            \sum_{m=1}^{\infty}
            \left[
             A_{im}^2
             + \frac{R}{\sqrt{3} a} A_{im} A_{i,m+1}
            \right] \, .
            \label{Eq:norm_A2}
        \end{align}

    \item[(iii)] $E$ representation:
        In this case the poles of the $S$-matrix are twofold
        degenerate~\cite{GasRic1989c}, such that our results do not immediately
        apply.
\end{enumerate}

We mention as an aside that we have numerically evaluated
\cref{Eq:norm_A2}
for the example $R/a = 2.1$ and $k_i a \approx 201.99 - 0.45 \ui$, which is
straightforward using the numerically determined coefficients $A_{im}$ of
the resonance state.
We find agreement when comparing to a direct Zel'dovich-regularized
two-dimensional volume computation, \cref{Eq:norm_regularized}.
Note that this volume integration is more challenging than the evaluation
of \cref{Eq:norm_A2}, as the result depends on numerical parameters like grid resolution,
cut-off of the integration domain, and implementation of the limiting procedure
$\eta \to 0^+$.

\subsection{Piecewise-constant potential}%
\label{sec:piecewise_constant_potential}

We now focus on scattering from a piecewise-constant potential
$V(\vecr)$ in $D$ dimensions.
For simplicity, the potential is assumed to have just two values,
namely
\begin{equation}
    V(\vecr) =
    \begin{cases}
        V_\Gamma \, , & \vecr \in \Gamma \\
        0 \, , & \text{otherwise}
    \end{cases}
    \quad .
    \label{Eq:piecewise_constant_potential}
\end{equation}
Here, resonance states $\psi(\vecr)$ and their gradients
$\nabla \psi(\vecr)$ are continuous across the boundary $\partial \Gamma$.
The case of more regions with different values of the potential can be obtained
by straightforward generalization of the following arguments.

\subsubsection{Norm}

We show that the norm $\mathcal{N}_i$ of a resonance state
can be computed by a boundary integral.
We split up the integral, \cref{Eq:norm_regularized}, into two regions,
the inner region $\Gamma$ and the outer region $\mathbb{R}^D \setminus \Gamma$.
We use the vector field $\RVec(\vecr)$, \cref{Eq:R_definition} with
$\lambda = k_i^2 - V_\Gamma$ for the inner and $\lambda = k_i^2$ for the outer
region and $\psi = \psi_i$.
It has the desired divergence
$\nabla \cdot \RVec(\vecr) = \left[\psi_i(\vecr) \right]^2$,
\cref{Eq:R_divergence}, in both regions.

For the integral in the inner region $\Gamma$ we immediately use the divergence
theorem.
For the integral in the outer region $\mathbb{R}^D \setminus \Gamma$ we
apply the divergence theorem according
to~\cref{Eq:regularized_integral_divergence} with $\boldsymbol{f}(\vecr) = \RVec(\vecr)$.
Together this yields
\begin{equation}
    \mathcal{N}_i
    =
    \oint_{\partial \Gamma} \dS \
    \nuVec(\vecr) \cdot \left[ \RVec^{\text{in}}(\vecr) - \RVec^{\text{out}}(\vecr) \right] \, ,
    \label{Eq:norm_piecewise_constant_R_difference}
\end{equation}
where the superscripts `in' and `out'
of $\RVec(\vecr)$ indicate whether it is evaluated on the inner or the outer
side of the boundary~$\partial \Gamma$.
For the following discussion it is convenient to write $\RVec(\vecr)$,
\cref{Eq:R_definition},
with $\lambda = k_i^2 - V(\vecr)$ for both regions,
\begin{equation}
    \RVec(\vecr) = \frac{1}{2} \, \vecr \, \psi_i\psi_i
    + \frac{1}{2} \frac{1}{k_i^2 - V(\vecr)} \,
    \Big[
    \left(D - 2 \right) \psi_i \, \nabla \psi_i
    - \vecr \, \left( \nabla \psi_i \cdot \nabla \psi_i \right)
    + 2 (\vecr \cdot \nabla \psi_i) \, \nabla \psi_i
    \Big]
    \label{Eq:R_definition_potential}
    \, ,
\end{equation}
where $\psi_i$ and $\nabla \psi_i$ are continuous across the boundary,
while the factor $k_i^2 - V(\vecr)$ is discontinuous.
Thus, only the first term of $\RVec(\vecr)$ cancels when evaluating the difference
in \cref{Eq:norm_piecewise_constant_R_difference} and we find
\begin{align}
    \mathcal{N}_i
    =
    \frac{V_\Gamma}{2 \, k_i^2 (k_i^2 - V_\Gamma)}
    \oint_{\partial \Gamma} \dS \
    \Big\{
    & \left[ \nuVec(\vecr) \cdot \nabla \psi_i(\vecr) \right]
    \left[
    \left(D - 2 \right) \psi_i(\vecr)
    + 2\vecr \cdot \nabla \psi_i(\vecr)
    \right]
    \nonumber \\
    &
    -\left[ \nuVec(\vecr) \cdot \vecr \right]
    \left[ \nabla \psi_i(\vecr) \cdot \nabla \psi_i(\vecr) \right]
    \Big\} \, .
    \label{Eq:norm_piecewise_constant}
\end{align}
Using that the gradient can be decomposed into a normal and a tangential component,
\cref{Eq:nabla_normal_tangential}, we can express the norm as
\begin{align}
    \mathcal{N}_i
    =
    \frac{V_\Gamma}{2 \, k_i^2 (k_i^2 - V_\Gamma)}
    \oint_{\partial \Gamma} \dS \
    \Big\{
    & \partial_\nu \psi_i(\vecr) \left[
    \left(D - 2 \right) \psi_i(\vecr)
    + 2\vecr \cdot \nabla_{\tVec} \psi_i(\vecr)
    \right]
    \nonumber \\
    &
    - \left[ \nuVec(\vecr) \cdot \vecr \right]
    \left[ \nabla_{\tVec} \psi_i(\vecr) \cdot \nabla_{\tVec}
        \psi_i(\vecr)
          - \left\{ \partial_{\nu} \psi_i(\vecr) \right\}^2 \right]
    \Big\} \, .
    \label{Eq:norm_piecewise_constant_tangential}
\end{align}
This result reduces to the case of a hard-wall potential,
\cref{Eq:norm_hard_wall},
in the limit $V_\Gamma \to \infty$ by using that
$\psi_i(\vecr) = 0$, \cref{Eq:boundary_conditions_wf_obstacle},
and $\nabla_{\tVec} \psi_i(\vecr) = 0$,
\cref{Eq:gradient_obstacle}, on the boundary $\partial \Gamma$.

\subsubsection{Example: One-dimensional potential}%
\label{sec:one_dimensional_potential}

We illustrate the evaluation of the boundary integral for the norm,
\cref{Eq:norm_piecewise_constant}, for the two
simplest one-dimensional piecewise-constant potentials and the general
piecewise-constant case, all with a hard-wall potential for $x < 0$.

\begin{itemize}
\item[(i)] First, we consider a paradigmatic example for scattering in one
dimension~\cite{Moi2011}, a constant step potential $V_\Gamma$ in region
$\Gamma = [0, a]$,
\begin{equation}
    V(x) =
    \begin{cases}
        \infty \, , & x < 0 \\
        V_\Gamma \, , & x \in [0, a] \\
        0 \, , & x > a
    \end{cases}
    \quad .
\end{equation}
Let us remark that
without the hard-wall potential, i.e., with zero potential for $x < 0$,
one could immediately apply~\cref{Eq:norm_piecewise_constant} with $D = 1$
for the boundary $\partial \Gamma$ consisting of two points, $x = 0$ and $x = a$.
With the hard-wall potential, for $x = 0$ there is no contribution from $\RVec^{\text{out}}(\vecr)$
in~\cref{Eq:norm_piecewise_constant_R_difference}.
The contribution from $\RVec^{\text{in}}(\vecr)$ vanishes,
as we have $\psi(0) = 0$ (due to the hard-wall potential)
and $\vecr = 0$ (due to the convenient choice of the origin of the coordinate system).
Thus there is no contribution from $x = 0$ to the norm.
For $x = a$, we have $\nuVec \cdot \vecr = a$ and $\nabla = \partial_x$,
such that we find
\begin{equation}
    \mathcal{N}_i
    =
    \frac{V_\Gamma}{2 k_i^2 (k_i^2 - V_\Gamma)} \,
    \partial_{x} \psi_i(a)
    \left[ - \psi_i(a) + a \, \partial_x \psi_i(a) \right] \, .
\end{equation}

Specifically, if one writes the solution as
\begin{equation}
    \psi_i(x) =
    \begin{cases}
        \sin\tilde{k}_i x \, , & 0 \le x \leq a \\
        C_i \, \exp[\ui k_i (x-a)] \, , & x > a
    \end{cases}
    \quad ,
\end{equation}
with $\tilde{k}_i = \sqrt{k_i^2 - V_\Gamma}$ and
$C_i = \sin\tilde{k}_i a$
such that $\psi_i(x)$ is continuous at $x=a$,
then we find the norm
\begin{equation}
    \mathcal{N}_i
    =
    \frac{1}{2} \left(a + \frac{\ui}{k_i}\right)
    \, .
    \label{Eq:norm_piecewise_constant_step}
\end{equation}
This agrees with the norm from integration, see Ref.~\cite[Eq.~(6.114)]{Moi2011}.
In the derivation of this result and for the comparison we use the continuity of the first
derivative leading to
\begin{align}
    \frac{\psi_i(a)}{\partial_x \psi_i(a)} &= - \frac{\ui}{k_i}
    \, , \\
    \left[ \partial_x \psi_i(a) \right]^2
    &= \tilde{k}_i^2 \cos^2 \tilde{k}_i a
    = \frac{\tilde{k}_i^2}{1 + \tan^2 \tilde{k}_i a}
    = \frac{\tilde{k}_i^2}{1 + \tilde{k}_i^2
          \left[\frac{\psi_i(a)}{\partial_x \psi_i(a)}\right]^2}
    = \frac{k_i^2 \left(k_i^2 - V_\Gamma\right)}{V_\Gamma}
    \, .
\end{align}
Note that the case $k_i = 0$ does not occur for resonance
states in this system.

\item[(ii)] Second, we consider a barrier potential with constant potential $V_\Gamma$ in
region $\Gamma = [a, b]$~\cite{Lee2009},
\begin{equation}
    V(x) =
    \begin{cases}
        \infty \, , & x < 0 \\
        0 \, , & 0 \le x < a \\
        V_\Gamma \, , & a \le x < b \\
        0 \, , & x \ge b
    \end{cases}
    \quad .
\end{equation}
The contribution at the hard wall vanishes, see discussion above.
We find for the norm
\begin{equation}
    \mathcal{N}_i =
    \frac{V_\Gamma}{2 k_i^2 (k_i^2 - V_\Gamma)} \,
    \left\{
        \partial_{x} \psi_i(b)
        \left[ - \psi_i(b) + b \, \partial_x \psi_i(b) \right]
        -
        \partial_{x} \psi_i(a)
        \left[ - \psi_i(a) + a \, \partial_x \psi_i(a) \right]
    \right\} \, .
\end{equation}
In comparison to the result from Ref.~\cite[Eq.~(16)]{Lee2009} this expression
for the norm does not
involve an integral over the inner region $x \in [0, b]$, but just the values of
the resonance state and its derivative at the two boundary points $x = a$ and
$x = b$.
Note that the cases $k_i=0$ and $k_i^2 = V_\Gamma$ do not
occur for resonance states in this system.

\item[(iii)] More generally, we consider a one-dimensional potential with
$n$ regions $[x_l, x_{l+1}]$ of piecewise-constant potential~$V_l$ for $l=0,\dots,n-1$
and $V_n=0$ for $x >x_n$,
\begin{equation}
    V(x) =
    \begin{cases}
        \infty \, , & x < x_0 \\
        V_l \, , & x_l \le x < x_{l+1}  \qquad l=0,\dots,n-1\\
        0 \, , & x \ge x_n
    \end{cases}
    \quad .
\end{equation}
Choosing $x_0=0$ is convenient, as then the contribution at the hard wall vanishes,
see discussion above. We find the norm by
generalizing the derivation of~\cref{Eq:norm_piecewise_constant} for
$D = 1$,
\begin{equation}
    \mathcal{N}_i =
    \sum_{l=1}^{n}
    \frac{V_{l-1} - V_l}{2 (k_i^2 - V_l) (k_i^2 - V_{l-1})} \,
        \partial_{x} \psi_i(x_l)
        \left[ - \psi_i(x_l) + x_l \, \partial_x \psi_i(x_l) \right]
    \, .
\end{equation}
Note that the case $k_i^2 = V_l$ does not
occur for resonance states in this system.
\end{itemize}

\subsection{Arbitrary potential}%
\label{sec:smooth_potential}

Finally, we consider scattering from an arbitrary smooth potential
$V(\vecr)$ in $D$ dimensions.
We express the integrand of \cref{Eq:norm_regularized} using the divergence,
\cref{Eq:R_divergence_potential}, of the vector field $\RVec(\vecr)$,
\cref{Eq:R_definition}, with $\lambda = k_i^2$ and $\psi = \psi_i$,
\begin{align}
    \left[\psi_i(\vecr) \right]^2
    =
    \nabla \cdot \RVec(\vecr)
    - \frac{V(\vecr) \psi_i(\vecr)}{2k_i^2}
    \Big[ \left(D - 2\right) \psi_i(\vecr)  + 2 \vecr \cdot \nabla \psi_i(\vecr) \Big]
    \label{Eq:R_divergence_smooth_potential}
    \, .
\end{align}

When evaluating the norm, \cref{Eq:norm_regularized}, there are now two
contributions corresponding to the two terms in \cref{Eq:R_divergence_smooth_potential}.
For the first term we apply the divergence theorem using
\cref{Eq:regularized_integral_divergence_no_Gamma} with
$\boldsymbol{f}(\vecr)=\RVec(\vecr)$, such that the contribution vanishes.
The second contribution comes from the potential term
in~\cref{Eq:R_divergence_smooth_potential} and directly leads to the norm
\begin{align}
    \mathcal{N}_i
    =
    \frac{1}{2k_i^2} \,
    \int_{\mathbb{R}^D} \dV \
    V(\vecr) \psi_i(\vecr)
    \Big[ \left(2 - D \right) \psi_i(\vecr)
    - 2 \vecr \cdot \nabla \psi_i(\vecr) \Big]
    \label{Eq:norm_smooth}
    \, .
\end{align}
In general this integral needs to be regularized as in
\cref{Eq:norm_regularized} and cannot be reduced to a boundary integral.
Depending on the type of potential $V(\vecr)$, however, compared to
\cref{Eq:norm_regularized} it can be an advantage that the integrand is
proportional to the potential $V(\vecr)$:
\begin{itemize}
\item[(i)] For potentials $V(\vecr)$ that are zero outside some region $\Gamma$, the
integral in \cref{Eq:norm_smooth} can be restricted to~$\Gamma$.

\item[(ii)] For potentials $V(\vecr)$ that decay faster than exponentially for
$r \to \infty$, the exponential growth of the resonance states is no
longer a problem and the integral can be computed in the main scattering
region only.
\end{itemize}

\newpage

\section{Electromagnetic systems}%
\label{sec:electromagnetic_systems}

Resonance states in electromagnetic systems correspond to time-harmonic electric
and magnetic fields,
$\EVec(\vecr, t) = \EVec(\vecr) \ue^{-\ui \omega t}$ and
$\HVec(\vecr, t) = \HVec(\vecr) \ue^{-\ui \omega t}$,
which are solutions to the source-free Maxwell equations.
In the following we need the two curl equations,
\begin{align}
    \nabla \times \EVec(\vecr) & = \ui \omega \, \mutens(\vecr, \omega) \, \HVec(\vecr) \, ,
    \label{Eq:maxwell3}
    \\
    \nabla \times \HVec(\vecr) & = - \ui \omega \,  \epstens(\vecr, \omega) \, \EVec(\vecr) \, ,
    \label{Eq:maxwell4}
\end{align}
where $\epstens(\vecr, \omega)$ and $\mutens(\vecr, \omega)$ are the
permittivity and permeability tensors, respectively, which in general depend on
the position $\vecr$ and the complex-valued frequency $\omega$.
Further, for $r \to \infty$ resonance states satisfy the outgoing wave boundary
condition, \cref{Eq:asymptotic_form}, with the complex wavenumber
$k = \omega \sqrt{\varepsilon_0 \mu_0}$ in vacuum.
This is fulfilled for discrete resonance frequencies $\omega_i$ and corresponding
fields $\EVec_i(\vecr)$ and $\HVec_i(\vecr)$, which are the resonance states of the system.
We here restrict ourselves to reciprocal materials for which
$\epstens^\top = \epstens$ and $\mutens^\top = \mutens$.
The case of bi-anisotropic materials can be included~\cite{MulWei2018}, but is
left out for clarity.

In~\cref{sec:norm_biorthogonality_electromagnetic_systems} we give a short
derivation of the established norm and biorthogonality of resonance states in
electromagnetic systems.
For piecewise-homogeneous materials in three dimensions we derive a boundary integral expression for
the norm in~\cref{sec:norm_piecewise_homogeneous_electromagnetic_systems}
and apply it to the example of a spherical scatterer (Mie scattering).
In~\cref{sec:2d_cavities}, which is independent of \cref{sec:norm_piecewise_homogeneous_electromagnetic_systems}, we consider two-dimensional cavities and derive a
one-dimensional boundary integral expression for the norm, which is applied to
the example of a circular disk.

\subsection{Norm and biorthogonality}%
\label{sec:norm_biorthogonality_electromagnetic_systems}
\subsubsection{Norm}
\label{sec:normalization_electromagnetic_systems}

The normalization of resonance states in electromagnetic systems has been established over the last two decades,
see the review~\cite{SauWuZarMulLal2022} and references therein.
For systems embedded
in vacuum the norm~$\mathcal{N}_i$ of the $i$-th resonance state
can be evaluated by~\cite{MulWei2018,StoColBonMcP2021,SauWuZarMulLal2022}
\begin{align}
    \mathcal{N}_i = &
    \int_{\Omega} \dV \
    \left\{
        \EVec_i \cdot \partial_\omega\big[\omega \epstens(\vecr, \omega)\big]_{\omega_i} \EVec_i
        - \HVec_i \cdot \partial_\omega\big[\omega \mutens(\vecr, \omega)\big]_{\omega_i} \HVec_i
    \right\}
    \nonumber \\
    & +
    \frac{\ui}{\omega_i} \oint_{\partial \Omega} \dS \ \nuVec \cdot \left[
        \EVec_i \times \left( \vecr \cdot \nabla \right) \HVec_i
        + \HVec_i \times \left( \vecr \cdot \nabla \right) \EVec_i
    \right] \, ,
    \label{Eq:norm_electrodynamics}
\end{align}
where $\Omega$ is a volume enclosing the system including all inhomogeneities,
$\nuVec$ is the outward-pointing unit normal vector on the surface
$\partial \Omega$, and the derivatives with respect to $\omega$ are assumed to exist and are taken at
$\omega = \omega_i$.
In the boundary integral on the second line, the fields are evaluated on the
outer side of the surface $\partial \Omega$.

Note that in this paper we use SI units for the fields $\EVec$ and $\HVec$.
Consequently the norm~$\mathcal{N}_i$ has the units of energy.
In the literature on normalization often other units are chosen, such that
$\EVec$ and $\HVec$ have the same unit and the norm is dimensionless.
This allows for choosing a normalization such that the norm is $1$~\cite{SauWuZarMulLal2022}
or $k_i R$~\cite{StoColBonMcP2021}.
Using SI units a reasonable normalization
of the fields might be that their norm is
given by the energy of a photon at the
resonance frequency, i.e., a complex-valued norm
$\mathcal{N}_i = \hbar \omega_i$
or a real-valued norm
$\mathcal{N}_i = \hbar \, \text{Re} \,\omega_i$.
We leave this choice to the reader and restrict ourselves to the determination
of the norm~$\mathcal{N}_i$ of a numerically or analytically given resonance
state.

In the following we give a short and transparent derivation of
the norm, \cref{Eq:norm_electrodynamics}, from biorthogonality in nonlinear
eigenvalue problems, \cref{Eq:nonlinear_biorthogonality}.
Here, the operator is given by the Maxwell equations.
The use of Zel'dovich regularization in the electromagnetic context
is similar to Ref.~\cite{StoColBonMcP2021}.
Specifically, the two Maxwell equations, \cref{Eq:maxwell3,Eq:maxwell4}, can be
written as a matrix equation
\begin{equation}
    T(\omega) \, \psiR(\vecr) = 0 \, ,
\end{equation}
with the combined field vector
\begin{equation}
    \psiR(\vecr) = \begin{pmatrix}
    \EVec(\vecr) \\
    \HVec(\vecr)
    \end{pmatrix} \, ,
\end{equation}
and the $6 \times 6$ matrix operator
\begin{equation}
    T(\omega) = \begin{pmatrix}
    \omega \, \epstens(\vecr, \omega) & - \ui \nabla \times \\
    - \ui \nabla \times & - \omega \, \mutens(\vecr, \omega)
    \end{pmatrix} \, .
    \label{Eq:definition_T_electrodynamics}
\end{equation}
Note that often the combined field vector $\psiR$ is defined with a prefactor of $\ui$
for the magnetic field, $(\EVec, \ui \HVec)$, which is just as well
possible.
The equations are written such that the operator is symmetric,
$T^\top(\omega) = T(\omega)$, where we use for reciprocal materials that
the permittivity and permeability tensors are symmetric
as well as the symmetry of the curl operator, $(\nabla \times)^{\top} = \nabla \times$
(which is due to the $3 \times 3$ matrix in Cartesian coordinates being antisymmetric and
each spatial derivative being antisymmetric).
As a consequence, the left eigenvector $\psiL(\vecr)$ is simply the complex
conjugate of the right eigenvector $\psiR(\vecr)$,
see~\cref{Eq:left_right_relation}.

According to~\cref{Eq:nonlinear_norm_regularized} the norm of a resonance state
is given in position space by using Zel'dovich regularization as
\begin{align}
    \mathcal{N}_i & =
    \langle \psiL_i | T'(\omega_i) | \psiR_i \rangle \nonumber \\
    & =
    \lim_{\eta \to 0^+} \int_{\mathbb{R}^3} \dV \ \psiR_i(\vecr) \cdot T'(\omega_i) \, \psiR_i(\vecr) \ \ue^{-\eta r^2} \nonumber \\
    & =
    \lim_{\eta \to 0^+}\int_{\mathbb{R}^3} \dV
    \left\{
        \EVec_i \cdot \partial_\omega\big[\omega \epstens(\vecr, \omega)\big]_{\omega_i} \EVec_i
        - \HVec_i \cdot \partial_\omega\big[\omega \mutens(\vecr, \omega)\big]_{\omega_i} \HVec_i
    \right\}
    \ue^{-\eta r^2} \, .
    \label{Eq:norm_electrodynamics_volume}
\end{align}
In this form it is impractical to evaluate the norm, since the integral is over
all space.
In order to avoid this, one splits up the integral into
an inner region $\Omega$ that
contains all inhomogeneities and the outer region
$\mathbb{R}^3 \setminus \Omega$ as a first step.
The inner integral leads to the first line in~\cref{Eq:norm_electrodynamics} for the norm.
For the integral in the outer region we use the vacuum properties
$\epstens(\vecr, \omega) = \varepsilon_0 \mathbb{1}$ and
$\mutens(\vecr, \omega) = \mu_0 \mathbb{1}$.
In this region the integrand can be written as a divergence,
\begin{equation}
    \varepsilon_0 \EVec_i(\vecr) \cdot \EVec_i(\vecr)
    - \mu_0 \HVec_i(\vecr) \cdot  \HVec_i(\vecr)
    = \nabla \cdot \MVec_i(\vecr)
    \label{Eq:integrand_as_divergence_electrodynamics}
\end{equation}
with the vector field~\cite{MulWei2018, StoColBonMcP2021}
\begin{align}
    \MVec_i(\vecr)
    = &
    - \frac{\ui}{\omega_i}
    \left[
        \EVec_i(\vecr) \times \left( \vecr \cdot \nabla \right)
        \HVec_i(\vecr)
        + \HVec_i(\vecr) \times \left( \vecr \cdot \nabla \right)
        \EVec_i(\vecr)
    \right] \, ,
    \label{Eq:vectorfield_electrodynamics}
\end{align}
which can be verified using vector calculus and Maxwell equations in vacuum.
Applying the divergence theorem with Zel'dovich regularization for the integral
in the outer region, according to \cref{Eq:regularized_integral_divergence} with
$\boldsymbol{f}(\vecr) = \MVec_i(\vecr)$ one ends up with
the surface integral in the second line in~\cref{Eq:norm_electrodynamics} for the norm.
This concludes the short derivation of~\cref{Eq:norm_electrodynamics} for the norm
with everything specific to electromagnetic systems given in the above derivation.

\subsubsection{Biorthogonality}
\label{sec:biorthogonality_electromagnetic_systems}

The biorthogonality relation between resonance states $i \neq j$
for nonlinear operators is given
by~\cref{Eq:nonlinear_biorthogonality,Eq:nonlinear_biorthogonality_regularized}
in position space as
\begin{align}
    0
    & = \langle \psiL_i | \frac{T(\omega_i) - T(\omega_j)}{\omega_i - \omega_j} | \psiR_j \rangle \nonumber \\
    & =
    \lim_{\eta \to 0^+} \int_{\mathbb{R}^3} \dV \ \psiR_i(\vecr) \cdot
    \left[ \frac{T(\omega_i) - T(\omega_j)}{\omega_i - \omega_j} \right] \,
    \psiR_j(\vecr) \ \ue^{-\eta r^2} \nonumber \\
    & =
    \frac{1}{\omega_i - \omega_j} \
    \lim_{\eta \to 0^+} \int_{\mathbb{R}^3} \dV \ g_{ij}(\vecr) \ \ue^{-\eta r^2} \, ,
    \label{Eq:generalized_biorthogonality_offdiagonal_position}
\end{align}
with the scalar field
\begin{align}
    g_{ij}(\vecr) =
    \EVec_i(\vecr) \cdot \left[ \omega_i \epstens(\vecr, \omega_i) - \omega_j \epstens(\vecr, \omega_j) \right] \EVec_j(\vecr)
    - \HVec_i(\vecr) \cdot \left[ \omega_i \mutens(\vecr, \omega_i) - \omega_j \mutens(\vecr, \omega_j) \right] \HVec_j(\vecr)
\end{align}
where we used $T(\omega)$ from \cref{Eq:definition_T_electrodynamics}.
It was already shown in Ref.~\cite{StoColBonMcP2021} that the integral indeed
vanishes using Zel'dovich regularization.
For convenience of the reader we quickly repeat the argument:
the scalar field $g_{ij}(\vecr)$ can be written as a divergence,
\begin{equation}
    g_{ij}(\vecr) = \nabla \cdot \GVec_{ij}(\vecr)
    \quad \text{with} \quad
    \GVec_{ij}(\vecr) =
    \ui
    \left[
    \EVec_i(\vecr) \times \HVec_j(\vecr)
    - \EVec_j(\vecr) \times \HVec_i(\vecr)
    \right] \, ,
\end{equation}
which can be verified by using the Maxwell equations for a reciprocal material,
\cref{Eq:maxwell3,Eq:maxwell4}, and vector calculus.
Applying the divergence theorem with Zel'dovich regularization to \cref{Eq:generalized_biorthogonality_offdiagonal_position},
according to \cref{Eq:regularized_integral_divergence_no_Gamma}
with $\boldsymbol{f}(\vecr) = \GVec_{ij}(\vecr)$
and $\nuVec \cdot \GVec_{ij}$ continuous across the boundary,
one finds indeed that the integral vanishes.
This demonstrates biorthogonality of left and right resonance states with respect to the generalized biorthogonal scalar product~\cref{Eq:nonlinear_biorthogonality,Eq:nonlinear_biorthogonality_regularized}.

\subsection{Norm for piecewise-homogeneous materials}%
\label{sec:norm_piecewise_homogeneous_electromagnetic_systems}
If the material is homogeneous within a finite region $\Gamma$,
possibly consisting of several unconnected regions, \cref{Eq:region_gamma},
we show in the following that the norm of a resonance state
can be expressed by a single boundary integral.
In particular, contributions that are continuous across the border cancel each
other out.

This approach is of numerical relevance, as cancellations in the norm
from the volume and the surface term in
\cref{Eq:norm_electrodynamics}
are automatically considered, reducing possible numerical errors.
Additionally, such a boundary integral needs less grid points for numerical
evaluation compared to the volume integral appearing in
\cref{Eq:norm_electrodynamics}.
Furthermore, if resonance states are originally determined on the boundary,
as e.g.\ in the boundary element method for two-dimensional
cavities~\cite{Wie2003},
the determination of the norm is particularly fast and accurate.

In the following we start with the simplest case of an
isotropic and non-dispersive material, i.e., a material whose permittivity and
permeability do not depend on frequency (in the considered frequency range).
Afterwards we give the norm for a material that is either anisotropic or dispersive,
while the most general case of anisotropic and dispersive material remains a challenge.
Generalization to systems consisting of several piecewise
homogeneous materials with different properties is straightforward.

We start from~\cref{Eq:norm_electrodynamics} for the norm and
choose $\Omega = \Gamma$ to be the region of homogeneous material
with boundary~$\partial \Gamma$.
For better readability we drop the index $i$ in the following,
with $\mathcal{N}$, $\EVec$, $\HVec$, $\MVec$, $\omega$, and $k$
referring to the $i$-th resonance state.

\subsubsection{Isotropic and non-dispersive material}

If the material inside the region $\Gamma$ is
isotropic and non-dispersive, we there have
$\epstens(\vecr, \omega) = \varepsilon_0 \epsrel \mathbb{1}$ and
$\mutens(\vecr, \omega) = \mu_0 \murel \mathbb{1}$, with relative
permittivity $\epsrel$ and relative permeability $\murel$.
Then the integrand in the first line of~\cref{Eq:norm_electrodynamics}
is the divergence of
the vector field $\MVec(\vecr)$ defined in~\cref{Eq:vectorfield_electrodynamics}.
The integrand in the second line of~\cref{Eq:norm_electrodynamics} is given by
the vector field $\MVec(\vecr)$ evaluated on the outer side
of~$\partial \Gamma$.
We thus find
\begin{align}
    \mathcal{N} = &
    \oint_{\partial \Gamma} \dS \ \nuVec(\vecr) \cdot
    \left[
        \MVec^{\text{in}}(\vecr) - \MVec^{\text{out}}(\vecr)
    \right] \, ,
    \label{Eq:norm_electrodynamics_boundary_M_difference}
\end{align}
where $\MVec^{\text{in}}(\vecr)$ and $\MVec^{\text{out}}(\vecr)$ denote
the vector field $\MVec(\vecr)$, \cref{Eq:vectorfield_electrodynamics},
evaluated at the inner and outer side of the boundary, respectively.
Note that this expression is similar to the case of a quantum particle in a piecewise-constant potential,
\cref{Eq:norm_piecewise_constant_R_difference}, but with a different vector
field.

Further simplification is achieved by canceling terms, which are continuous
across the boundary.
To this end we decompose the fields on the boundary $\partial \Gamma$
into a tangential and a normal component,
$\EVec = \EVec_{\tVec} + \nuVec E_\nu$ and
$\HVec = \HVec_{\tVec} + \nuVec H_\nu$.
This decomposition is not possible at discontinuities of a
piecewise-smooth boundary.
However, such discontinuities can be excluded
from the boundary integral
\cref{Eq:norm_electrodynamics_boundary_M_difference}, as the integrand is
bounded.
Therefore, in the rest of
\cref{sec:norm_piecewise_homogeneous_electromagnetic_systems} the boundary
integral $\oint_{\partial \Gamma}$ excludes all discontinuities.
The decomposition of the fields and the product rule leads for the scalar
product of $\nuVec \cdot \MVec$ to
\begin{align}
    \nuVec \cdot \MVec
    &=
        - \frac{\ui}{\omega}
        \nuVec \cdot
        \Big[
            \EVec_{\tVec} \times
            \left( \vecr \cdot \nabla \right) \HVec_{\tVec}
            + H_\nu \EVec_{\tVec} \times
            \left( \vecr \cdot \nabla \right) \nuVec
            + \HVec_{\tVec} \times
            \left( \vecr \cdot \nabla \right) \EVec_{\tVec}
            + E_\nu \HVec_{\tVec} \times
            \left( \vecr \cdot \nabla \right) \nuVec
        \Big]
    \nonumber \\
    &=
    - \frac{\ui}{\omega}
    \nuVec \cdot
    \Big\{
        \EVec_{\tVec} \times
        \left[
            \left(\nuVec \cdot \vecr \right) \partial_\nu \HVec_{\tVec}
            + (\vecr \cdot \nabla_{\tVec}) \HVec_{\tVec}
        \right]
        + \HVec_{\tVec} \times
        \left[
            \left(\nuVec \cdot \vecr \right) \partial_\nu \EVec_{\tVec}
            + (\vecr \cdot \nabla_{\tVec}) \EVec_{\tVec}
        \right]
    \Big\}
    \nonumber \\
    &
    \hspace*{0.4cm} + \frac{\ui}{\omega}
    \left[
        H_\nu \EVec_{\tVec}
        + E_\nu \HVec_{\tVec}
    \right]
    \cdot
    \left[
        \nuVec \times (\vecr \cdot \nabla_{\tVec}) \nuVec
    \right]
    \, ,
\end{align}
where in the second step we use the decomposition of the gradient,
\cref{Eq:nabla_normal_tangential},
$\nabla = \nuVec \partial_\nu + \nabla_{\tVec}$, $\partial_\nu \nuVec = 0$
(choice of extension of $\nuVec$ away from the boundary),
and the permutation property of the scalar triple product.
Using that $\EVec_{\tVec}$, $\HVec_{\tVec}$,
and their tangential derivatives
are continuous across the boundary,
we find for the integrand in~\cref{Eq:norm_electrodynamics_boundary_M_difference},
\begin{align}
    \nuVec \cdot
    \left[
        \MVec^{\text{in}} - \MVec^{\text{out}}
    \right]
    =
    &- \frac{\ui}{\omega} \left[ \nuVec \cdot \vecr \right]
    \nuVec \cdot
    \Big\{
        \EVec_{\tVec} \times
        \left[
            \partial_\nu \HVec_{\tVec}^{\text{in}}
            - \partial_\nu \HVec_{\tVec}^{\text{out}}
        \right]
        + \HVec_{\tVec} \times
        \left[
            \partial_\nu \EVec_{\tVec}^{\text{in}}
            - \partial_\nu \EVec_{\tVec}^{\text{out}}
        \right]
    \Big\}
    \nonumber \\
    &
    + \frac{\ui}{\omega}
        \left\{
            \left[ H_\nu^{\text{in}} - H_\nu^{\text{out}} \right]
            \EVec_{\tVec}
            + \left[ E_\nu^{\text{in}} - E_\nu^{\text{out}} \right]
            \HVec_{\tVec}
        \right\}
        \cdot
        \left[
            \nuVec \times (\vecr \cdot \nabla_{\tVec}) \nuVec
        \right]
    \, .
    \label{Eq:norm_electrodynamics_boundary_nu_dot_M_difference}
\end{align}

Furthermore, it is possible to express the integrand just by the
fields inside (or outside) the region~$\Gamma$.
For this we use the properties of the normal components for isotropic materials,
$E^{\text{out}}_\nu = \epsrel E^{\text{in}}_\nu$ and
$H^{\text{out}}_\nu = \murel H^{\text{in}}_\nu$.
They extend to the tangential derivative,
$\nabla_{\tVec} E^{\text{out}}_\nu = \epsrel \nabla_{\tVec} E^{\text{in}}_\nu$ and
$\nabla_{\tVec} H^{\text{out}}_\nu = \murel \nabla_{\tVec} H^{\text{in}}_\nu$.
It remains to relate this tangential derivative of the normal component of both fields
to the normal derivative of the tangential component,
$\partial_\nu \EVec_{\tVec}$ and $\partial_\nu \HVec_{\tVec}$,
appearing in~\cref{Eq:norm_electrodynamics_boundary_nu_dot_M_difference}.
This is achieved by multiplying the two
Maxwell equations, \cref{Eq:maxwell3,Eq:maxwell4}, from the left by $\nuVec \times$,
using the formula for the gradient of a dot product for $\nabla(\nuVec \cdot \EVec)$
and $\nabla(\nuVec \cdot \HVec)$ with $\nabla \times \nuVec = 0$, and taking the tangential component of the equations,
    \begin{align}
        \partial_\nu \EVec_{\tVec}
        & =
        \nabla_{\tVec} E_\nu - (\EVec_{\tVec} \cdot \nabla_{\tVec}) \nuVec - \ui \omega \mu(\vecr) \nuVec \times \HVec_{\tVec}
        \, ,
        \label{Eq:boundary_condition_normal_derivative_E}
        \\
        \partial_\nu \HVec_{\tVec}
        & =
        \nabla_{\tVec} H_\nu - (\HVec_{\tVec} \cdot \nabla_{\tVec}) \nuVec + \ui \omega \varepsilon(\vecr) \nuVec \times \EVec_{\tVec}
                        \, .
        \label{Eq:boundary_condition_normal_derivative_H}
    \end{align}
When inserting \cref{Eq:boundary_condition_normal_derivative_E,Eq:boundary_condition_normal_derivative_H} into \cref{Eq:norm_electrodynamics_boundary_nu_dot_M_difference}, the terms
$ (\EVec_{\tVec} \cdot \nabla_{\tVec}) \nuVec$ and $(\HVec_{\tVec} \cdot \nabla_{\tVec}) \nuVec$ cancel, as they are continuous across the boundary.
Using additionally that
$\nuVec \cdot [\EVec_{\tVec} \times (\nuVec \times \EVec_{\tVec})] = \EVec_{\tVec} \cdot \EVec_{\tVec}$ and
$\nuVec \cdot [\HVec_{\tVec} \times (\nuVec \times \HVec_{\tVec})] = \HVec_{\tVec} \cdot \HVec_{\tVec}$
gives for the norm
\begin{align}
    \mathcal{N} =
    \oint_{\partial \Gamma} \dS \
    &\left[ \nuVec(\vecr) \cdot \vecr \right] \,
    \begin{aligned}[t]
        \Bigg(
            &\hphantom{+} \hspace{0.2cm} (\epsrel - 1) \,
            \left\{
                \varepsilon_0 \EVec_{\tVec}(\vecr) \cdot \EVec_{\tVec}(\vecr)
                + \frac{\ui}{\omega} \nuVec(\vecr) \cdot
                \left[
                    \HVec_{\tVec}(\vecr) \times \nabla_{\tVec} E^{\, \text{in}}_{\nu}(\vecr)
                \right]
            \right\}
            \nonumber \\
            &+ (\murel - 1) \,
            \left\{
                - \mu_0 \HVec_{\tVec}(\vecr) \cdot \HVec_{\tVec}(\vecr)
                + \frac{\ui}{\omega} \nuVec(\vecr) \cdot
                \left[
                    \EVec_{\tVec}(\vecr) \times \nabla_{\tVec} H^{\, \text{in}}_{\nu}(\vecr)
                \right]
            \right\}
        \Bigg)
    \end{aligned}
    \nonumber \\
    &
    - \frac{\ui}{\omega}
    \left[
        (\murel - 1) H_\nu^{\, \text{in}}(\vecr) \EVec_{\tVec}(\vecr)
        + (\epsrel - 1) E_\nu^{\, \text{in}}(\vecr) \HVec_{\tVec}(\vecr)
    \right]
    \cdot
    \left[
        \nuVec(\vecr) \times (\vecr \cdot \nabla_{\tVec}) \nuVec(\vecr)
    \right]
    \, .
    \label{Eq:norm_electrodynamics_boundary}
\end{align}
This expression for the norm is just a boundary integral.
Here it is expressed by the normal component of both fields on the inner
side of the boundary,
but it could just as well be expressed by the value on the outer side of the
boundary.

The expression for the norm in~\cref{Eq:norm_electrodynamics_boundary}
is valid for a piecewise-smooth boundary $\partial \Gamma$, excluding all
discontinuities. For a smooth
boundary $\partial \Gamma$ further simplification is possible, see
\cref{sec:appendix_smooth_boundary}, and one finds
\begin{align}
    \mathcal{N}
    =\oint_{\partial \Gamma} \dS \
    \left[ \nuVec(\vecr) \cdot \vecr \right] \,
    \Bigg(
        &\hphantom{+} \hspace{0.2cm}
        (\epsrel - 1) \, \varepsilon_0
        \left\{
            \EVec_{\tVec}(\vecr) \cdot \EVec_{\tVec}(\vecr)
            + \epsrel \left[ E_\nu^{\, \text{in}}(\vecr) \right]^2
        \right\}
        \nonumber \\
        &- (\murel - 1) \, \mu_0
        \left\{
            \HVec_{\tVec}(\vecr) \cdot \HVec_{\tVec}(\vecr)
            + \murel \left[ H_\nu^{\, \text{in}}(\vecr) \right]^2
        \right\}
    \Bigg)
    \, ,
    \label{Eq:norm_electrodynamics_boundary_smooth}
\end{align}
which is easier to evaluate, as it contains no derivatives of the fields.

\subsubsection{Anisotropic and non-dispersive material}%
\label{sec:normalization_electromagnetic_systems_anisotropic}

If the material in region~$\Gamma$ is anisotropic (but reciprocal) and non-dispersive
most of the above reasoning is unchanged.
In particular, the vector field $\MVec(\vecr)$, \cref{Eq:vectorfield_electrodynamics}, can still be used since its divergence
gives the integrand in~\cref{Eq:norm_electrodynamics_volume} for the norm
when using $(\vecr \cdot \nabla) \epstens \EVec = \epstens (\vecr \cdot \nabla) \EVec$
and $(\vecr \cdot \nabla) \mutens \HVec = \mutens (\vecr \cdot \nabla) \HVec$
for piecewise constant $\epstens$ and $\mutens$.
Thus, \cref{Eq:norm_electrodynamics_boundary_M_difference,Eq:norm_electrodynamics_boundary_nu_dot_M_difference} can still be used
to compute the norm.
If one wants to express the integrand just by fields inside (or outside) of the region~$\Gamma$,
one would have to take the boundary conditions in anisotropic materials into account.

\subsubsection{Isotropic and dispersive material}

If the material in region $\Gamma$ is isotropic and dispersive,
$\epstens(\vecr, \omega) = \varepsilon_0 \epsrel(\omega) \mathbb{1}$ and
$\mutens(\vecr, \omega) = \mu_0 \murel(\omega) \mathbb{1}$,
we introduce a vector field
\begin{align}
    \MVecTilde(\vecr)
    = &
    \,\MVec(\vecr)
    - \frac{\ui}{2}
    \left\{
    \left(\frac{\varepsilon'}{\varepsilon} + \frac{\mu'}{\mu} \right)
    \left[
        \EVec(\vecr) \times \left( \vecr \cdot \nabla \right) \HVec(\vecr)
        + \HVec(\vecr) \times \left( \vecr \cdot \nabla \right) \EVec(\vecr)
    \right]
    +
    \left(\frac{\varepsilon'}{\varepsilon} - \frac{\mu'}{\mu} \right)
    \left[
        \EVec(\vecr) \times \HVec(\vecr)
    \right]
    \right\}
    \, ,
    \label{Eq:vectorfield_electrodynamics_frequency_dependent}
\end{align}
with $\MVec(\vecr)$ from \cref{Eq:vectorfield_electrodynamics} and the
abbreviations $\varepsilon' = \partial_\omega \varepsilon(\omega)$ and
$\mu' = \partial_\omega \mu(\omega)$,
such that its divergence
\begin{equation}
    \nabla \cdot \MVecTilde(\vecr)
    =
    \partial_\omega\big[\omega \varepsilon(\omega)\big] \EVec(\vecr) \cdot \EVec(\vecr)
    - \partial_\omega\big[\omega \mu(\omega)\big] \HVec(\vecr) \cdot  \HVec(\vecr)
    \,
\end{equation}
gives the integrand in the first line of \cref{Eq:norm_electrodynamics}.
Using that $\MVecTilde^{\text{out}}(\vecr) = \MVec^{\text{out}}(\vecr)$
and the decomposition of the fields into tangential and normal components
this leads to the norm
\begin{align}
    \tilde{\mathcal{N}} = &
    \oint_{\partial \Gamma} \dS \ \nuVec(\vecr) \cdot
    \left[
        \MVecTilde^{\text{in}}(\vecr) - \MVec^{\text{out}}(\vecr)
    \right]
    \nonumber
    \\
    = &
    \,\mathcal{N}
    - \frac{\ui}{2}
    \oint_{\partial \Gamma} \dS \
    \bigg\{ \
        \hphantom{+}
        \left(\frac{\epsrel'}{\epsrel} + \frac{\murel'}{\murel} \right)
        \nuVec(\vecr) \cdot
        \left[
            \EVec_{\tVec}(\vecr) \times \left( \vecr \cdot \nabla \right) \HVec_{\tVec}^{\text{in}}(\vecr)
            + \HVec_{\tVec}(\vecr) \times \left( \vecr \cdot \nabla \right) \EVec_{\tVec}^{\text{in}}(\vecr)
        \right]
        \nonumber \\
        & \hspace*{2.7cm}
        -
        \left(\frac{\epsrel'}{\epsrel} + \frac{\murel'}{\murel} \right)
        \left[
            H^\text{in}_\nu(\vecr) \EVec_{\tVec}(\vecr)
            + E^\text{in}_\nu(\vecr) \HVec_{\tVec}(\vecr)
        \right]
        \cdot
        \left[
            \nuVec(\vecr) \times (\vecr \cdot \nabla_{\tVec}) \nuVec(\vecr)
        \right]
        \nonumber \\
        & \hspace*{2.7cm} +
        \left(\frac{\epsrel'}{\epsrel} - \frac{\murel'}{\murel} \right)
        \nuVec(\vecr) \cdot
        \left[
            \EVec_{\tVec}(\vecr) \times \HVec_{\tVec}(\vecr)
        \right]
    \bigg\}
    \, ,
    \label{Eq:norm_electrodynamics_boundary_frequency_dependent}
\end{align}
with $\mathcal{N}$ from \cref{Eq:norm_electrodynamics_boundary}
with $\varepsilon(\omega) = \varepsilon_0 \epsrel(\omega)$
and $\mu(\omega) = \mu_0 \murel(\omega)$
as well as the abbreviations
$\epsrel' = \partial_\omega \epsrel(\omega)$ and
$\murel' = \partial_\omega \murel(\omega)$.

\subsubsection{Example: Spherical scatterer}

For the example of an isotropic three-dimensional spherical scatterer of radius
$R$ with homogeneous relative permittivity $\epsrel(\omega)$ and permeability
$\murel(\omega)$, we compute the norm according to
\cref{Eq:norm_electrodynamics_boundary_smooth,Eq:norm_electrodynamics_boundary_frequency_dependent}
for non-dispersive and dispersive materials, respectively.
Compared to previous calculations this is much simpler, as we only need to
evaluate a boundary integral on the surface of the sphere, which is directly given by
the normalization of the spherical harmonics.
We find corrections by a factor of $2$ compared to previous
results~\cite{DooLanMul2014,MulLan2016,StoColBonMcP2021},
see detailed discussion at the end of this section.

For resonance states, the electric and magnetic field inside
of the sphere ($r \leq R$) are given
for TE modes (also called odd modes or magnetic type)
by~\cite{Str2015}
\begin{align}
    \EVec^\mathrm{TE}(\vecr) &=
    E_0 \boldsymbol{\psi}(\vecr)
    \, ,
    \label{Eq:sphere_E_TE}
    \\
    \HVec^\mathrm{TE}(\vecr) &=
    -\frac{\ui}{\omega \mu_0 \murel} \nabla \times \EVec^\mathrm{TE}(\vecr)
    = -\frac{\ui \varepsilon_0 E_0 \omega }{\murel k^2 r} \boldsymbol{\chi}(\vecr)
    \, ,
\end{align}
and for TM modes (also called even modes or electric type) by~\cite{Str2015}
\begin{align}
    \HVec^\mathrm{TM}(\vecr) &=
    H_0 \boldsymbol{\psi}(\vecr)
    \, ,
    \label{Eq:sphere_H_TM}
    \\
    \EVec^\mathrm{TM}(\vecr) &=
    \frac{\ui}{\omega \varepsilon_0 \epsrel} \nabla \times \HVec^\mathrm{TM}(\vecr)
    = \frac{\ui \mu_0 H_0 \omega}{\epsrel k^2 r} \boldsymbol{\chi}(\vecr)
    \, ,
\end{align}
where in both cases the second field is explicitly fulfilling the Maxwell equations,
\cref{Eq:maxwell3,Eq:maxwell4},
and we used $\omega^2 = k^2 /  (\varepsilon_0 \mu_0)$.
The fields outside of the sphere ($r > R$) are not needed for evaluating the norm according
to
\cref{Eq:norm_electrodynamics_boundary_smooth,Eq:norm_electrodynamics_boundary_frequency_dependent}.
The dimensionless vector fields $\boldsymbol{\psi}(\vecr)$ and
$\boldsymbol{\chi}(\vecr)$ are given in the spherical basis
$\{\hat{\vecr}, \hat{\boldsymbol{\theta}}, \hat{\boldsymbol{\varphi}}\}$
by
\begin{align}
    \boldsymbol{\psi}(\vecr) &=
    \rho(r)
    \begin{pmatrix}
        0 \\
        \frac{1}{\sin \theta} \partial_\varphi Y_{lm}(\Omega) \\
        - \partial_\theta Y_{lm}(\Omega)
    \end{pmatrix}
    \, ,
    \\
    \boldsymbol{\chi}(\vecr) &=
    r \, \nabla \times \boldsymbol{\psi}(\vecr) =
    \begin{pmatrix}
        \rho(r) \, l (l + 1) Y_{lm}(\Omega) \\
        \left[ \partial_r r \rho(r) \right]
        \partial_\theta Y_{lm}(\Omega) \\
        \left[ \partial_r r \rho(r) \right]
        \frac{1}{\sin \theta} \partial_\varphi Y_{lm}(\Omega)
    \end{pmatrix}
    \, ,
    \label{Eq:sphere_chi}
\end{align}
with $l \in \mathbb{N}$ and $m = -l, \ldots, l$.
The angular dependence is given by the real (tesseral) spherical harmonics
$Y_{lm}(\Omega)$, which are defined such that
\begin{equation}
    \int \ud \Omega \, \ Y_{lm}(\Omega) \, Y_{l'm'}(\Omega) =
    \delta_{ll'} \delta_{mm'} \, ,
    \label{Eq:spherical_harmonics_orthonormality}
\end{equation}
and are explicitly given in Ref.~\cite{DooLanMul2014}.
They are eigenfunctions of the angular part of the Laplacian on the unit sphere,
\begin{equation}
    \Delta_\Omega Y_{lm}(\Omega)
    = \frac{1}{\sin \theta} \partial_\theta
    \left( \sin \theta \, \partial_\theta Y_{lm}(\Omega) \right)
    + \frac{1}{\sin^2 \theta} \partial_{\varphi \varphi} Y_{lm}(\Omega)
    = - l (l + 1) Y_{lm}(\Omega)
    \, .
    \label{Eq:spherical_harmonics_eigenvalue_equation}
\end{equation}
The radial dependence for $r \leq R$ is given by
\begin{equation}
    \rho(r) = \frac{j_l(\nref k r)}{j_l(\nref k R)} \, ,
\end{equation}
with $\nref = \sqrt{\epsrel \murel}$ the refractive index and $j_l(z)$ the
spherical Bessel function of order $l$
fulfilling~\cite[Eq.~(\href{https://dlmf.nist.gov/10.47.E1}{10.47.1})]{DLMF},
\begin{equation}
    \left[
        z^2 \partial_{zz} + 2 z \partial_z +
        z^2 - l (l + 1)
    \right] j_l(z) = 0 \, .
    \label{Eq:spherical_bessel_equation}
\end{equation}

For non-dispersive materials,
\cref{Eq:norm_electrodynamics_boundary_smooth}, we use that on the sphere ($r = R$),
\begin{align}
    \nuVec \cdot \vecr &= R \, ,
    \\
    \EVec_{\tVec} \cdot \EVec_{\tVec}
    &=
    E_\theta^2 + E_\varphi^2 \, ,
    \\
    E_{\nu}^{\,\text{in}}
    &=
    E_r
    \, ,
    \\
    \HVec_{\tVec} \cdot \HVec_{\tVec}
    &=
    H_\theta^2 + H_\varphi^2 \, ,
    \\
    H_{\nu}^{\,\text{in}}
    &=
    H_r
    \, .
\end{align}
For dispersive materials,
\cref{Eq:norm_electrodynamics_boundary_frequency_dependent},
we use additionally that,
\begin{align}
    \nuVec \cdot [\EVec_{\tVec} \times
    (\vecr \cdot \nabla) \HVec_{\tVec}^{\text{in}}]
    &=
    E_\theta R \, \partial_r H_\varphi
    - E_\varphi R \, \partial_r H_\theta \, ,
    \\
    \nuVec \cdot [\HVec_{\tVec} \times
    (\vecr \cdot \nabla) \EVec_{\tVec}^{\text{in}}]
    &=
    H_\theta R \, \partial_r E_\varphi
    - H_\varphi R \, \partial_r E_\theta \, ,
    \\
    \nuVec \cdot [\EVec_{\tVec} \times
    \HVec_{\tVec}]
    &= E_\theta H_\varphi - E_\varphi H_\theta \, ,
    \\
    \vecr \cdot \nabla_{\tVec} &= 0
    \, .
\end{align}
Furthermore, we define for each resonance state a dimensionless constant $\alpha$
for a term appearing in $\boldsymbol{\chi}(\vecr)$, \cref{Eq:sphere_chi},
\begin{align}
    \alpha
    = \partial_r r \rho(r) \Big\rvert_{r=R}
    = \rho(R) + R \, \partial_r \rho(r)
    \Big\rvert_{r=R}
    = 1 + \nref k R \frac{j'_l(\nref k R)}{j_l(\nref k R)}
    \, ,
    \label{Eq:alpha}
\end{align}
where we use $\rho(R) = 1$.
Additionally, we use \cref{Eq:spherical_bessel_equation} to
evaluate
\begin{equation}
    R \, \partial_{rr} r \rho(r) \Big\rvert_{r=R}
    = l (l + 1) - \nref^2 k^2 R^2
    \, .
\end{equation}
When evaluating the norm according to
\cref{Eq:norm_electrodynamics_boundary_smooth,Eq:norm_electrodynamics_boundary_frequency_dependent}
for all terms the same boundary integral appears,
\begin{align}
    \oint_{\partial \Gamma} \dS \, \left[ \nabla_\Omega Y_{lm}(\Omega) \right]^2
    &= - R^2 \int \ud \Omega \, Y_{lm}(\Omega) \Delta_\Omega Y_{lm}(\Omega)
    \nonumber \\
    &= l (l + 1) R^2 \int \ud \Omega \, \left[Y_{lm}(\Omega)\right]^2
    = l (l + 1) R^2 \, ,
\end{align}
where we use Green's first identity and
the properties of the real spherical harmonics,
\cref{Eq:spherical_harmonics_eigenvalue_equation,Eq:spherical_harmonics_orthonormality}.
So one just has to consider the appropriate factors at $r=R$.

For non-dispersive materials,
\cref{Eq:norm_electrodynamics_boundary_smooth}, we find for the norm of TE and TM modes
\begin{align}
    \mathcal{N}^\mathrm{\, TE} &=
    \phantom{-}
    l \left(l + 1\right)
    \left\{
        (\epsrel - 1)
        +
        \frac{\murel - 1}{\murel k^2 R^2}
        \left[
            l (l + 1)
            + \frac{\alpha^2}{\murel}
        \right]
    \right\}
    R^3 \varepsilon_0 E_0^2 \, ,
    \label{Eq:sphere_norm_TE_nondispersive}
    \\
    \mathcal{N}^\mathrm{\, TM} &=
    - l \left(l + 1\right)
    \left\{
        (\murel - 1)
        +
        \frac{\epsrel - 1}{\epsrel k^2 R^2}
        \left[
            l (l + 1)
            + \frac{\alpha^2}{\epsrel}
        \right]
    \right\}
    R^3 \mu_0 H_0^2 \, .
    \label{Eq:sphere_norm_TM_nondispersive}
\end{align}
For dispersive materials,
\cref{Eq:norm_electrodynamics_boundary_frequency_dependent},
we find
\begin{align}
    \tilde{\mathcal{N}}^\mathrm{\, TE} &=
    \mathcal{N}^\mathrm{\, TE}
    + \frac{l \left(l + 1\right) \omega}{2 \murel k^2 R^2}
    \left\{
        \left(\frac{\epsrel'}{\epsrel} + \frac{\murel'}{\murel}\right)
        \left[
            \alpha^2
            + \nref^2 k^2 R^2
            - l (l + 1)
        \right]
        +
        \left(\frac{\murel'}{\murel} - \frac{\epsrel'}{\epsrel} \right)
        \alpha
    \right\}
    R^3 \varepsilon_0 E_0^2
    \, ,
    \label{Eq:sphere_norm_TE_dispersive}
    \\
    \tilde{\mathcal{N}}^\mathrm{\, TM} &=
    \mathcal{N}^\mathrm{\, TM}
    - \frac{l \left(l + 1\right) \omega}{2 \epsrel k^2 R^2}
    \left\{
        \left(\frac{\epsrel'}{\epsrel} + \frac{\murel'}{\murel}\right)
        \left[
            \alpha^2
            + \nref^2 k^2 R^2
            - l (l + 1)
        \right]
        +
        \left(\frac{\epsrel'}{\epsrel} - \frac{\murel'}{\murel}\right)
        \alpha
    \right\}
    R^3 \mu_0 H_0^2
    \, .
    \label{Eq:sphere_norm_TM_dispersive}
\end{align}
In the above equations we separate the dimensionless factors from the
factors $R^3 \varepsilon_0 E_0^2$  and $R^3 \mu_0 H_0^2$, which have units of energy.
The constant $\alpha$ for each resonance state is defined in~\cref{Eq:alpha}.

We find agreement with Refs.~\cite{DooLanMul2014, MulLan2016} up to conventions,
and a correction by a factor of $2$ for Ref.~\cite{StoColBonMcP2021}.
We first choose $E_0$ for TE modes in~\cref{Eq:sphere_E_TE} and $H_0$
for TM modes in~\cref{Eq:sphere_H_TM}
such that our definition of the resonance state is identical
to the one in the literature.
Then we determine the norm according to the above
\cref{Eq:sphere_norm_TE_nondispersive,Eq:sphere_norm_TM_nondispersive,Eq:sphere_norm_TE_dispersive,Eq:sphere_norm_TM_dispersive}
and check if the desired value~$1$~\cite{DooLanMul2014, MulLan2016} or
$kR$~\cite{StoColBonMcP2021} is found.
Note that in these references different units for electric and magnetic fields
are used as well as $\varepsilon_0 = \mu_0 = 1$.

\begin{itemize}
\item[(i)] For comparison with Ref.~\cite{DooLanMul2014} for $\murel = 1$ and
non-dispersive materials we use
$E_0 = A^\mathrm{TE}_l$ and $H_0 = -\ui A^\mathrm{TM}_l$ with
$A^\mathrm{TE/TM}_l$ from Ref.~\cite[Eq.~(29)]{DooLanMul2014},
yielding $\mathcal{N}^\mathrm{\, TE} = \mathcal{N}^\mathrm{\, TM} = 2$.
For comparison with Ref.~\cite{MulLan2016} for $\murel = 1$ and dispersive materials we use
$H_0 = -\ui A^\mathrm{TM}_l$ with
$A^\mathrm{TM}_l$ from Ref.~\cite[Eq.~(D5)]{MulLan2016},
yielding $\tilde{\mathcal{N}}^\mathrm{\, TM} = 2$.
These results are consistent as the definition of the norm in
Refs.~\cite{DooLanMul2014, MulLan2016} differs by a factor of two compared to
\cref{Eq:norm_electrodynamics}, see discussion in Ref.~\cite{MulWei2018}.

\item[(ii)] For comparison with Ref.~\cite{StoColBonMcP2021} for the general case
$\murel \neq 1$ and dispersive materials we use for TE modes
$E_0 = \frac{1}{\mathcal{N}_\alpha} h_l(kR) k^{3/2} \ui^l \frac{1}{\sqrt{2}} \frac{1}{\sqrt{l(l+1)}}$
with $\mathcal{N}_\alpha$ from Ref.~\cite[Eq.~(37a)]{StoColBonMcP2021}
and $h_l(z)$ the spherical Hankel function of the first kind of order $l$,
yielding $\mathcal{N}^\mathrm{\, TE} = (-1)^l \frac{1}{2} kR$.
The first factor $(-1)^l$ was explicitly neglected in
Ref.~\cite[App.~F]{StoColBonMcP2021}, the second factor $\frac{1}{2}$ is due to
an inadvertent oversight in the angular
integration in Ref.~\cite[App.~F]{StoColBonMcP2021}
related to the chosen normalization of the real spherical harmonics,
and the third factor $kR$ is the desired normalization chosen in Ref.~\cite{StoColBonMcP2021}.
The comparison for TM modes leads to the same overall agreement.
\end{itemize}

Note that in this example we have a $(2l+1)$-fold degeneracy.
Even though we have excluded such cases from our general considerations, we find
that in this example the norm can still be computed as the chosen basis in the degenerate subspace is biorthogonal.
Choosing a different basis, e.g., complex spherical harmonics, would instead lead to a vanishing norm and no biorthogonality.

\subsection{Two-dimensional cavities}%
\label{sec:2d_cavities}

Deformed disk cavities provide a simple yet powerful platform for
studying wave chaos phenomena with applications ranging from photonic devices
to fundamental investigations of mode dynamics and chaos in optical
systems~\cite{CaoWie2015, JiaChaZhaWieHenWanCaoXiaAlu2026}.
They are modeled as a deformed cylinder with infinite extension and
translational symmetry in $z$-direction.
Therefore, the electric and magnetic fields can be written as
$\EVec(\vecr) = \EVec(x,y) \, \ue^{\ui k_z z}$ and
$\HVec(\vecr) = \HVec(x,y) \, \ue^{\ui k_z z}$, where $k_z$ is the
wavenumber in $z$-direction.
We consider the case $k_z=0$, such that the system is effectively
two-dimensional,
namely the electric and magnetic fields
depend on the $x$- and $y$-coordinate only.
For clarity we restrict ourselves to isotropic and non-dispersive materials, i.e.,
$\epstens(\vecr, \omega) = \varepsilon_0 \, \epsrel(\vecr) \mathbb{1}$
and $\mutens(\vecr, \omega) = \mu_0 \, \murel(\vecr) \mathbb{1}$.
Generalization to anisotropic or dispersive materials is possible.
The relative permittivity $\epsrel(\vecr)$ and relative permeability $\murel(\vecr)$
are assumed to be constant functions within a region $\Gamma$, \cref{Eq:region_gamma},
that extends in $z$ direction and is embedded in vacuum,
\begin{align}
    \epsrel(\vecr) &= \begin{cases}
                    \epsrel \, , \quad \vecr \in \Gamma \\
                    1 \, , ~\, \quad \text{otherwise}
                    \end{cases}  ,
    \label{Eq:definition_epsrel_3D}
    \\
    \murel(\vecr) &= \begin{cases}
                    \murel \, , \quad \vecr \in \Gamma \\
                    1 \, , ~\, \quad \text{otherwise}
                    \end{cases} .
    \label{Eq:definition_murel_3D}
\end{align}
In this case, the two Maxwell equations,~\cref{Eq:maxwell3,Eq:maxwell4},
decouple into two independent sets
of equations, for $(E_z, H_x, H_y)$ the transverse magnetic (TM)
polarization and for $(H_z, E_x, E_y)$ the transverse electric
(TE) polarization.

In the following we present the
case of TM polarization and provide the result for the norm for TE
polarization below.
The Maxwell equations,
\cref{Eq:maxwell3,Eq:maxwell4},
reduce for the non-zero field components
$(E_z, H_x, H_y)$ of TM polarization to
\begin{align}
    \partial_y E_z
    &=
    \ui \omega \mu_0 \murel(\vecr) H_x \, ,
    \label{Eq:maxwell_TM_x}
    \\
    -\partial_x E_z
    &=
    \ui \omega \mu_0 \murel(\vecr) H_y \, ,
    \label{Eq:maxwell_TM_y}
    \\
    \partial_x H_y - \partial_y H_x
    &=
    - \ui \omega \varepsilon_0 \epsrel(\vecr) E_z \, ,
    \label{Eq:maxwell_TM_z}
\end{align}
and can be further decoupled such that for the electric field $E_z$ one finds,
\begin{equation}
    \nabla \cdot \left( \frac{1}{\murel(\vecr)} \nabla E_z \right)
    =
    - \epsrel(\vecr) k^2 E_z
    \, ,
    \label{Eq:helmholtz_like_TM}
\end{equation}
with $k = \omega \sqrt{\varepsilon_0 \mu_0}$ the complex wavenumber in vacuum.

\subsubsection{Norm}

In the present geometry that extends to infinity in $z$-direction
one determines the norm per unit length in $z$-direction
starting from the integral in three-dimensional space,
\cref{Eq:norm_electrodynamics_volume},
and replacing it by a
two-dimensional integral in the infinite plane of constant $z$,
\begin{align}
    \mathcal{N}_\text{TM}
    & =
    \lim_{\eta \to 0^+}\int_{\mathbb{R}^2} \dV \
    \left\{
        \varepsilon_0 \epsrel(\vecr) E_z^2(\vecr)
        - \mu_0 \murel(\vecr) \left[H_x^2(\vecr) + H_y^2(\vecr)\right]
    \right\}
    \ue^{-\eta r^2}
    \, .
\end{align}
Here we use the properties of an isotropic, non-dispersive material and the
non-zero field components of TM polarization.
Note that one cannot apply the three-dimensional result,
\cref{Eq:norm_electrodynamics_boundary},
as it is derived for a finite region $\Gamma$ embedded in vacuum,
while here the region $\Gamma$ is assumed to extend to infinity.
Also, if one applies \cref{Eq:norm_electrodynamics_boundary} to a region
$\Gamma$ with finite height in $z$-direction (as in an experiment), the above
separation into TM and TE modes would not be exact and the fields have some
$z$-dependence due to corner effects.

Numerically, one often deals with the scalar electric field $E_z$ only and
the norm is expressed by
\begin{align}
    \mathcal{N}_\text{TM}
    &
    =
    \lim_{\eta \to 0^+}\int_{\mathbb{R}^2} \dV \
    \left\{
        2 \varepsilon_0 \epsrel(\vecr) E_z^2(\vecr)
        + \frac{\varepsilon_0}{k^2} \nabla \cdot \left[E_z(\vecr) \frac{\nabla E_z(\vecr)}{ \murel(\vecr)}\right]
    \right\}
    \ue^{-\eta r^2}
    \\
    &
    =
    2 \, \varepsilon_0
    \lim_{\eta \to 0^+}\int_{\mathbb{R}^2} \dV \
    \epsrel(\vecr) E_z^2(\vecr) \
    \ue^{-\eta r^2}
    \, ,
    \label{Eq:norm_TM_regularized}
\end{align}
where the first step can be verified using vector calculus and Maxwell equations for TM polarization,~\cref{Eq:maxwell_TM_x,Eq:maxwell_TM_y,Eq:maxwell_TM_z,Eq:helmholtz_like_TM}.
In the second step we apply the divergence theorem with Zel'dovich regularization to
the divergence term in the integrand according to
\cref{Eq:regularized_integral_divergence_no_Gamma}
with $\boldsymbol{f}(\vecr) = E_z(\vecr) \frac{\nabla E_z(\vecr)}{ \murel(\vecr)}$ and $\nuVec \cdot \boldsymbol{f}$ continuous across the boundary,
finding that its integral vanishes.
The factor 2 shows the equivalence of the contributions to the norm from the
electric field and the magnetic field, when integrated over the infinite plane.

Using the piecewise-constant material in region $\Gamma$,
\cref{Eq:definition_epsrel_3D,Eq:definition_murel_3D},
we split up~\cref{Eq:norm_TM_regularized} into
an integral over the inner region $\Gamma$
(from now on two-dimensional) and the outer region $\mathbb{R}^2 \setminus \Gamma$ of the cavity, and
find
\begin{align}
    \mathcal{N}_{\text{TM}} =
    2 \varepsilon_0\epsrel \int_{\Gamma} \dV \ E^2_z(\vecr)
    + 2 \varepsilon_0 \lim_{\eta \to 0^+} \int_{\mathbb{R}^2 \setminus \Gamma} \dV \ E^2_z(\vecr) \, \ue^{-\eta r^2}
    \, ,
\end{align}
where in the first term the limit $\eta \to 0^+$ has been taken.
We define a vector field $\RVec(\vecr)$, \cref{Eq:R_definition},
for both regions (excluding the boundary $\partial \Gamma$)
with $D=2$ and $\lambda = \epsrel(\vecr) \murel(\vecr) \, k^2 $,
\begin{align}
    \RVec(\vecr) &= \frac{1}{2} \, \vecr \, E_z^2
    + \frac{1}{2 \epsrel(\vecr) \murel(\vecr) \, k^2} \,
    \Big[
    - \vecr \, \left( \nabla E_z \cdot \nabla E_z \right)
    + 2 (\vecr \cdot \nabla E_z) \, \nabla E_z
    \Big]
    \, .
    \label{Eq:R_definition_electrodynamics}
\end{align}
It has the desired divergence
$\nabla \cdot \RVec(\vecr) = E^2_z(\vecr)$,
\cref{Eq:R_divergence},
in both regions.
Applying the divergence theorem to the first integral and
\cref{Eq:regularized_integral_divergence} to the second integral, where
$\boldsymbol{f}(\vecr) = \RVec(\vecr)$, we find that the norm can be expressed
as a one-dimensional boundary integral over $\partial \Gamma$,
\begin{equation}
    \mathcal{N}_{\text{TM}}
    =
    2 \varepsilon_0
    \oint_{\partial \Gamma} \ud s \
    \nuVec(s) \cdot
        \left[
        \epsrel \, \RVec^{\text{in}}(s) - \RVec^{\text{out}}(s)
        \right]
    \, ,
    \label{Eq:norm_boundary_TM_in_out}
\end{equation}
where $\RVec^{\text{in}}(s)$ and $\RVec^{\text{out}}(s)$ are
evaluated at arclength $s$ at the inner and outer side of the boundary, respectively.
Note that this expression is similar to the case of a quantum particle in a
piecewise-constant potential, \cref{Eq:norm_piecewise_constant_R_difference},
but due to the factor $\epsrel$ different terms cancel at the boundary.

Further simplification is achieved by canceling terms, which are continuous
across the boundary.
To this end, we split the gradient $\nabla E_z$ in~\cref{Eq:R_definition_electrodynamics}
into a normal component and a tangential component along the
boundary~$\partial \Gamma$,~\cref{Eq:nabla_normal_tangential}.
Furthermore, we replace the tangential derivative by a derivative with
respect to arclength $s$, $\nabla_{\tVec} = \tVec \partial_s$ with
unit tangent vector $\tVec = \hat{\boldsymbol{z}} \times \nuVec$ along the boundary,
and find
\begin{equation}
    \nuVec \cdot \RVec(\vecr) =
    \frac{1}{2} \left( \nuVec \cdot \vecr \right) \, E_z^2
    + \frac{1}{2 \epsrel(\vecr) \murel(\vecr) \, k^2} \,
    \Big\{
    - \left( \nuVec \cdot \vecr \right) \,
    \left[ (\partial_s E_z)^2 - (\partial_\nu E_z)^2 \right]
    + 2 \left(\vecr \cdot \tVec \right) \partial_s E_z \, \partial_{\nu} E_z
    \Big\}
    \, .
    \label{Eq:nu_dot_R_cavity}
\end{equation}
Here the factor $\epsrel(\vecr) \murel(\vecr)$
is discontinuous across the boundary,~\cref{Eq:definition_epsrel_3D,Eq:definition_murel_3D}.
The electric field $E_z$ is tangential (to the originally 3D region $\Gamma$) and thus continuous.
Therefore also its derivative along the boundary is continuous.
In contrast, it follows from \cref{Eq:maxwell_TM_x,Eq:maxwell_TM_y}
that $-\partial_{\nu} E_z(\vecr) = \ui \omega \, \mu_0 \, \murel(\vecr) H_{t}(\vecr)$. From the continuity of the tangential magnetic field $H_{t}(\vecr)$, it follows that $\frac{1}{\murel(\vecr)} \partial_{\nu} E_z(\vecr)$ is continuous across the boundary, leading to the
continuity relations
\begin{align}
    E_z^{\text{(in)}} & = E_z^{\text{(out)}} \, ,
    \label{Eq:boundary_condition_TM_Ez}\\
    \partial_{s} E_z^{\text{(in)}} & = \partial_{s} E_z^{\text{(out)}} \, ,
    \label{Eq:boundary_condition_TM_tangential_derivative}\\
    \partial_{\nu} E_z^{\text{(in)}} & = \murel \, \partial_{\nu} E_z^{\text{(out)}} \, .
    \label{Eq:boundary_condition_TM_normal_derivative}
\end{align}
We now cancel continuous contributions in \cref{Eq:norm_boundary_TM_in_out},
giving
\begin{align}
    \mathcal{N}_{\text{TM}}
    &=
    \varepsilon_0
    \oint_{\partial \Gamma} \ud s \
    \big[ \nuVec(s) \cdot \vecr(s) \big]
    \Big\{
        (\epsrel - 1) E_z^2(s)
        +
        \frac{\murel - 1}{\murel \, k^2}
            \Big(
        \left[ \partial_{s} E_z(s) \right]^2
        + \frac{1}{\murel} \left[ \partial_{\nu} E_z^{\, \text{in}}(s) \right]^2
        \Big)
    \Big\} \, .
    \label{Eq:norm_TM}
\end{align}
Here the discontinuous normal derivative $\partial_{\nu}$ is evaluated on the
inner side of the boundary, but using
\cref{Eq:boundary_condition_TM_normal_derivative} it
could just as well be expressed by the value on the outer side.

For TE polarization one finds that the result has the identical form
with $E_z$ replaced by $H_z$, $\varepsilon_0$ replaced by $\mu_0$,
and $\epsrel$ and $\murel$ interchanged, giving
\begin{align}
    \mathcal{N}_{\text{TE}}
    &=
    - \mu_0
    \oint_{\partial \Gamma} \ud s \
    \big[ \nuVec(s) \cdot \vecr(s) \big]
    \Big\{
        (\murel - 1) H_z^2(s)
        +
        \frac{\epsrel - 1}{\epsrel \, k^2}
            \Big(
        \left[ \partial_{s} H_z(s) \right]^2
        + \frac{1}{\epsrel} \left[ \partial_{\nu} H_z^{\, \text{in}}(s) \right]^2
        \Big)
    \Big\} \, .
    \label{Eq:norm_TE}
\end{align}
The minus sign has its origin in the minus sign of the magnetic field term in
the normalization, \cref{Eq:norm_electrodynamics_volume}.
For the special case of a dielectric cavity ($\murel = 1$),
some terms cancel in the above expressions.
Then, the norm is given
by~\cref{Eq:norm_TM_results,Eq:norm_TE_results} in \cref{sec:results} summarizing the
results.

We mention as an aside that we have numerically evaluated \cref{Eq:norm_TM}
for a dielectric cavity of lima\c{c}on shape with deformation parameter
$\epsilon = 0.6$~\cite{KetClaFriBae2022}, refractive index $\nref = 3.3$,
$\murel = 1$, reflection law for a TM polarized mode, and $k R_\text{cav}
\approx 10.23 - 0.03 \ui$ as shown in \cref{fig:overview}.
We find agreement when comparing to a direct Zel'dovich-regularized
two-dimensional volume computation, \cref{Eq:norm_TM_regularized}.
Note that this volume integration is more challenging than the boundary
integration, as the result depends on numerical parameters like grid resolution,
cut-off of the integration domain, and implementation of the limiting procedure
$\eta \to 0^+$.

\subsubsection{Biorthogonality}

Biorthogonality for pairs of resonance states $i \ne j$ holds also in the
present effectively two-dimensional setting, according to
\cref{Eq:generalized_biorthogonality_offdiagonal_position}
with the three-dimensional integral replaced by a two-dimensional one and
the electric and magnetic field components
corresponding to TM or TE polarization.
We show in addition, that biorthogonality also holds individually
for the electric field component $E_z$ (TM polarization),
\begin{align}
    0
    & =
    \lim_{\eta \to 0^+} \int_{\mathbb{R}^2} \dV \
    \epsrel(\vecr) \, E_{z,i}(\vecr) \, E_{z,j}(\vecr)
    \ \ue^{-\eta r^2}
    \,
    \label{Eq:biorthogonality_TM}
\end{align}
or the magnetic field component $H_z$ (TE polarization),
\begin{align}
    0
    & =
    \lim_{\eta \to 0^+} \int_{\mathbb{R}^2} \dV \
    \murel(\vecr) \, H_{z,i}(\vecr) \, H_{z,j}(\vecr)
    \ \ue^{-\eta r^2}
    \, .
    \label{Eq:biorthogonality_TE}
\end{align}
This is important, as numerically the most common case is to consider
such a scalar field only.

For the case of TM polarization we present the verification
by splitting the integral into
an integral over the inner and the outer region of the cavity
\begin{align}
    &
    \epsrel \int_{\Gamma} \dV \
    E_{z,i}(\vecr) \, E_{z,j}(\vecr) \
    + \lim_{\eta \to 0^+} \int_{\mathbb{R}^2 \setminus \Gamma} \dV \
    E_{z,i}(\vecr) \, E_{z,j}(\vecr) \, \ue^{-\eta r^2}
    \\
    & =
    \oint_{\partial \Gamma} \ud s \
    \underbrace{ \nuVec(s) \cdot
        \big[
        \epsrel \, \BVec_{ij}^{\text{in}}(s) - \BVec_{ij}^{\text{out}}(s)
        \big]
    }_{=0} = 0
    \, ,
\end{align}
where for the second line
we define a vector field $\BVec_{ij}(\vecr)$, \cref{Eq:L_definition},
with $\lambda_{i/j} = \epsrel \murel k_{i/j}^2$ in the inner region $\Gamma$
and $\lambda_{i/j} = k_{i/j}^2$ in the outer region $\mathbb{R}^2 \setminus \Gamma$
as well as $\psi_{i/j} = E_{z,i/j}$.
Its divergence in both regions is
$\nabla \cdot \BVec_{ij}(\vecr) = E_{z,i}(\vecr) \, E_{z,j}(\vecr)$,
\cref{Eq:L_divergence}.
Furthermore, we apply the divergence theorem to the integral in the inner region $\Gamma$
and with Zel'dovich regularization for the integral in the outer region
according to \cref{Eq:regularized_integral_divergence} with $\boldsymbol{f}(\vecr) = \BVec_{ij}^{\text{out}}(\vecr)$.
The integrand of the one-dimensional boundary integral over $\partial \Gamma$
vanishes everywhere, following from the boundary
conditions~\cref{Eq:boundary_condition_TM_Ez,Eq:boundary_condition_TM_normal_derivative},
confirming the biorthogonality relation, \cref{Eq:biorthogonality_TM}, for the
electric field component $E_z$.

\subsubsection{Example: Circular disk}

We compute the norm according to~\cref{Eq:norm_TM,Eq:norm_TE} for the example of
a circular disk cavity of radius $R$ and homogeneous refractive index.
For resonance states, the $z$-components of the electric and magnetic field
inside the cavity ($r \leq R$) for TM and TE polarization, respectively, are
given by~\cite{DetMorSieWaa2009,DooLanMul2013}
\begin{align}
    E_z^\mathrm{TM}(r,\varphi) &= E_0 \, \rho(r) \, \chi(\varphi)
    \, , \label{Eq:disk_TM_field}
    \\
    H_z^\mathrm{TE}(r,\varphi) &= H_0 \, \rho(r) \, \chi(\varphi) \, ,
\end{align}
where for $m \in \mathbb{Z}$ and resonance poles $k \in \mathbb{C}$ the
dimensionless functions $\rho$ and $\chi$ are defined as
\begin{equation}
    \rho(r) =
        \dfrac{J_m(\nref k r)}{J_m(\nref k R)} \quad
        \text{and} \quad \chi(\varphi) =
        \begin{cases}
            \frac{1}{\sqrt{\pi}} \sin m \varphi \, , & m < 0 \\
            \frac{1}{\sqrt{2\pi}} \, , & m=0 \\
            \frac{1}{\sqrt{\pi}} \, \cos m \varphi \, , &  m > 0
        \end{cases}
        \quad ,
\end{equation}
with normalization $\int_0^{2\pi} \ud \varphi \ \chi^2(\varphi) = 1$.
Here $\nref = \sqrt{\epsrel \murel}$ is the refractive index and $J_m$ is the
Bessel function of the first kind of order $m$.
The other field components ($H_x$, $H_y$ for TM polarization, $E_x$, $E_y$ for
TE polarization) can be computed from Maxwell equations, e.g.\
for TM polarization in~\cref{Eq:maxwell_TM_x,Eq:maxwell_TM_y}.

For TM polarization we use~\cref{Eq:norm_TM} with $\partial \Gamma$ being
the boundary of the disk, $\ud s = R \, \ud \varphi$,
$\nuVec \cdot \vecr = R$, $\partial_{s} = R^{-1} \, \partial_{\varphi}$, and
$\partial_{\nu} =  \partial_{r}$, giving for the norm
\begin{align}
    \mathcal{N}^{\, \text{TM}}
    &=
    \varepsilon_0 \, E_0^2
    \int_0^{2\pi} \ud \varphi \ R^2
    \Big\{
        (\epsrel - 1) \, \chi^2(\varphi)
        +
        \frac{\murel - 1}{\murel \, k^2}
            \Big(
        \left[\frac{1}{R} \, \chi'(\varphi) \right]^2
        + \epsrel \left[ k \, \frac{J'_m(\nref k R)}{J_m(\nref k R)} \, \chi(\varphi) \right]^2
        \Big)
    \Big\} \nonumber \\
    &=
    \left(
        (\epsrel - 1)
        +
        \frac{\murel - 1}{\murel}
            \left\{
        \frac{m^2}{k^2 R^2}
        + \epsrel \left[ \frac{J'_m(\nref k R)}{J_m(\nref k R)} \, \right]^2
        \right\}
    \right) R^2 \varepsilon_0 E_0^2
    \, .
    \label{Eq:norm_disk_TM}
\end{align}
For TE polarization we use~\cref{Eq:norm_TE} and analogously find
\begin{align}
    \mathcal{N}^{\, \text{TE}}
    &=
    -
    \left(
        (\murel - 1)
        +
        \frac{\epsrel - 1}{\epsrel}
            \left\{
        \frac{m^2}{k^2 R^2}
        + \murel \left[ \frac{J'_m(\nref k R)}{J_m(\nref k R)} \, \right]^2
        \right\}
    \right) R^2 \mu_0 H_0^2 \, .
    \label{Eq:norm_disk_TE}
\end{align}
In the above equations we separate the dimensionless factors from the
factors $R^2 \varepsilon_0 E_0^2$  and $R^2 \mu_0 H_0^2$, which have units of
energy per length.

We compare to the previous result~\cite{DooLanMul2013} for the
special case of a dielectric disk ($\murel = 1$) and TM polarization.
Note that in this reference different units for electric and magnetic fields
are used as well as $\varepsilon_0 = \mu_0 = 1$.
We first choose in~\cref{Eq:disk_TM_field} the factor $E_0 = A$ with $A$ from
Ref.~\cite[Eq.~(17)]{DooLanMul2013}, such that the
definition of the resonance state is identical to the one in
Ref.~\cite[Eq. (13)]{DooLanMul2013}.
Then we determine the norm $\mathcal{N}^{\, \text{TM}}$ according
to~\cref{Eq:norm_disk_TM} and find $\mathcal{N}^{\, \text{TM}}=2$.
This is consistent as the norm in Ref.~\cite{DooLanMul2013} contains the
electric field only and does not treat the (equally large) contribution from the
magnetic field, see discussion in Ref.~\cite{MulWei2018}.

Note that in this example we have a two-fold degeneracy ($\pm m$).
Even though we have excluded such cases from our general considerations, we find
that in this example the norm can still be computed as the chosen basis in the degenerate subspace is biorthogonal.
Choosing a different basis, e.g., $\ue^{\pm \ui m \varphi}$, would instead lead to a vanishing norm and no biorthogonality.

\newpage

\section{Summary}%
\label{sec:results}

In summary, we derive boundary integrals for computing the norm of resonance
states in quantum scattering systems and electromagnetic systems with piecewise
homogeneous properties bounded by piecewise-smooth boundaries.
This uses Zel'dovich regularization to deal with the exponential growth of the resonance
states.
Suitable vector fields allow for expressing the integrand of volume integrals as
a divergence, such that the divergence theorem can be applied giving boundary
integrals.
For the convenience of the reader, we here give a list of boundary integrals for
various systems derived in this paper.

For a quantum particle in various $D$-dimensional potentials, we find for the norm
in \cref{sec:potential}:
\begin{itemize}
    \item Hard-wall (infinite) potential in region $\Gamma$ (\cref{sec:hard_wall_potential}):
        \begin{align}
            \mathcal{N}_i
            &=
            -\frac{1}{2 k_i^2} \,
            \oint_{\partial \Gamma} \dS \
            \left[ \nuVec(\vecr) \cdot \vecr \right] \,
            \left[ \partial_{\nu} \psi_i(\vecr) \right]^2
            \label{Eq:norm_hard_wall_results}
        \end{align}
        As an example, we apply it to the three-disk scattering system in
        \cref{sec:three_disk_scattering} expressing the norm in terms of
        Fourier coefficients of the normal derivative of a resonance state
        along the boundary of the disks.

    \item Piecewise-constant potential with $V_\Gamma$ in region $\Gamma$ and
        zero potential outside (\cref{sec:piecewise_constant_potential}):
        \begin{align}
            \mathcal{N}_i
            =
            \frac{V_\Gamma}{2 \, k_i^2 (k_i^2 - V_\Gamma)}
            \oint_{\partial \Gamma} \dS \
            \Big\{
            & \left[ \nuVec(\vecr) \cdot \nabla \psi_i(\vecr) \right]
            \left[
            \left(D - 2 \right) \psi_i(\vecr)
            + 2\vecr \cdot \nabla \psi_i(\vecr)
            \right]
            \nonumber \\
            &
            -\left[ \nuVec(\vecr) \cdot \vecr \right]
            \left[ \nabla \psi_i(\vecr) \cdot \nabla \psi_i(\vecr) \right]
            \Big\}
            \label{Eq:norm_piecewise_constant_results}
        \end{align}
        Alternatively the gradient can be expressed in terms of the normal
        and tangential derivative along the boundary, see
        \cref{Eq:norm_piecewise_constant_tangential}.
        As an example, we apply it to hard-wall and piecewise-constant
        potentials in one dimension in \cref{sec:one_dimensional_potential}.

    \item Arbitrary, sufficiently fast decaying potential $V(\vecr)$
        (\cref{sec:smooth_potential}):
    \begin{align}
        \mathcal{N}_i
        =
        \frac{1}{2k_i^2} \,
        \int_{\mathbb{R}^D} \dV \
        V(\vecr) \psi_i(\vecr)
        \Big[ \left(2 - D \right) \psi_i(\vecr)
        - 2 \vecr \cdot \nabla \psi_i(\vecr) \Big]
        \label{Eq:norm_smooth_results}
    \end{align}
    This integral can be computed in the main scattering region only.
    Note that this case cannot be reduced to a boundary integral.
\end{itemize}

For resonance states of the Maxwell equations with a homogeneous,
isotropic, and non-dispersive material in region~$\Gamma$ and vacuum
outside, we find for the norm in \cref{sec:electromagnetic_systems}:
\begin{itemize}
    \item Three-dimensional region $\Gamma$
        (\cref{sec:norm_piecewise_homogeneous_electromagnetic_systems}):
    \begin{align}
        \mathcal{N}_i =
        \oint_{\partial \Gamma} \dS \
        &\left[ \nuVec(\vecr) \cdot \vecr \right] \,
        \begin{aligned}[t]
            \Bigg(
                &\hphantom{+} \hspace{0.2cm} (\epsrel - 1) \,
                \left\{
                    \varepsilon_0 \EVec_{\tVec,i}(\vecr) \cdot \EVec_{\tVec,i}(\vecr)
                    + \frac{\ui}{\omega_i} \nuVec(\vecr) \cdot
                    [\HVec_{\tVec,i}(\vecr) \times \nabla_{\tVec} E^{\, \text{in}}_{\nu,i}(\vecr)]
                \right\}
                \nonumber \\
                &+ (\murel - 1) \,
                \left\{
                    - \mu_0 \HVec_{\tVec,i}(\vecr) \cdot \HVec_{\tVec,i}(\vecr)
                    + \frac{\ui}{\omega_i} \nuVec(\vecr) \cdot
                    [\EVec_{\tVec,i}(\vecr) \times \nabla_{\tVec} H^{\, \text{in}}_{\nu,i}(\vecr)]
                \right\}
            \Bigg)
        \end{aligned}
        \nonumber \\
        &
        - \frac{\ui}{\omega_i}
        \left[
            (\murel - 1) H_{\nu,i}^{\, \text{in}}(\vecr) \EVec_{\tVec,i}(\vecr)
            + (\epsrel - 1) E_{\nu,i}^{\, \text{in}}(\vecr) \HVec_{\tVec,i}(\vecr)
        \right]
        \cdot
        \left[
            \nuVec(\vecr) \times (\vecr \cdot \nabla_{\tVec}) \nuVec(\vecr)
        \right]
        \, .
        \label{Eq:norm_electrodynamics_boundary_final}
    \end{align}
    If the boundary is smooth (instead of piecewise-smooth), a
    simplification is possible, see
    \cref{Eq:norm_electrodynamics_boundary_smooth}.
    The result for the more general case of dispersive materials is given by
    \cref{Eq:norm_electrodynamics_boundary_frequency_dependent} and materials
    that are anisotropic are discussed in
    \cref{sec:normalization_electromagnetic_systems_anisotropic}.
    The norm for the example of a spherical scatterer is given by
    \cref{Eq:sphere_norm_TE_dispersive,Eq:sphere_norm_TM_dispersive}
    for TE and TM modes, respectively.

    \item Two-dimensional dielectric cavity for TM and TE polarization
        (\cref{sec:2d_cavities}):
        \begin{align}
            \mathcal{N}_i^\mathrm{\,TM}
            & =
            \varepsilon_0 (\nrefsq -1)
            \oint_{\partial \Gamma} \ud s \
            \left[ \nuVec(s) \cdot \vecr(s) \right] \
            \left[ E_{z,i}(s) \right]^2 \, ,
            \label{Eq:norm_TM_results}
            \\
            \mathcal{N}_i^\mathrm{\,TE}
            & =
            - \frac{\mu_0(\nrefsq - 1)}{\nrefsq \, k_i^2}
            \oint_{\partial \Gamma} \ud s \
            \left[ \nuVec(s) \cdot \vecr(s) \right]
            \left\{
                \left[ \partial_{s} H_{z,i}(s) \right]^2
                + \frac{1}{\nrefsq}
                \big[ \partial_{\nu} H_{z,i}^{\text{in}}(s) \big]^2
            \right\}
            \label{Eq:norm_TE_results}
        \end{align}
        Here $\nref = \sqrt{\epsrel}$ is the refractive index inside the cavity.
        The result for materials that are in addition
        magnetic ($\murel \neq 1$) is given by
        \cref{Eq:norm_TM,Eq:norm_TE}.
        The norm for the example of a circular disk is given by
        \cref{Eq:norm_disk_TM,Eq:norm_disk_TE} for TM and TE polarization,
        respectively.
\end{itemize}

Generalization to systems with multiple domains of different homogeneous
properties is straightforward, see e.g., \cref{sec:one_dimensional_potential}(iii) for a
one-dimensional piecewise-constant potential.
Potentials for a quantum particle, which combine hard walls,
piecewise-constant parts, and arbitrary potential parts can be generalized from
these cases.
We are confident that these boundary integrals for the norm will provide a
useful tool, e.g., for computing the time evolution of wave packets in
scattering systems by using a large number of normalized resonance states.
Additionally, the methods presented here are useful to compute integrals
of resonance states over finite regions, e.g., for the phase
rigidity~\cite{LanBroBee1997, SadBer2005, RotBir2015, Wie2023, YiRyuRodHen2025}
or the Petermann factor~\cite{Sch2009b, Wie2023, KulWie2025}, by a boundary
integral.
For the characterization of resonance states of chaotic scattering systems,
their average near some decay rate is relevant~\cite{HarShi2015, KulWie2016,
BitKimZenWanCao2020, KetClaFriBae2022, KetLorSch2025}.
This requires adequate normalization, which can be achieved by the boundary
integrals derived here even in the semiclassical limit, i.e., for small
wavelengths compared to the size of the scattering system.
Furthermore, generalizing the presented normalization of resonance states to
degeneracies in the spectrum, e.g., exceptional points, is an interesting
direction for future research.

%%%%%%%%%%%%%%%%%%%%%%%%%%%%%%%%%%%%%%%%%%%%%%%%%%%%%%%%%%%%%%%%%%%%%%%%%%%%%
\acknowledgments

We are grateful for discussions with Stefan Bittner, Barbara Dietz, Takahisa Harayama, Martina Hentschel,
Maximilian Kieler, and Jan Wiersig.
We thank an anonymous referee for pointing out a missing contribution in
\cref{Eq:norm_electrodynamics_boundary_final}.
Funded by the Deutsche Forschungsgemeinschaft (DFG, German Research
Foundation)---262765445.
The authors gratefully acknowledge the computing time on the high-performance computer at the NHR Center of TU Dresden. This center is jointly supported by the Federal Ministry of Research, Technology and Space of Germany and the state governments participating in the NHR.

%%%%%%%%%%%%%%%%%%%%%%%%%%%%%%%%%%%%%%%%%%%%%%%%%%%%%%%%%%%%%%%%%%%%%%%%%%%%%
\newpage

\appendix

\section{Divergence of vector fields}%
\label{sec:appendix_divergence}

In this appendix we show the calculation of the divergence of vector fields
in $D$ dimensions defined in~\cref{sec:divergence},
using Cartesian coordinates with Einstein summation and $\partial_n x_n = D$.

The divergence of $\RVec(\vecr)$, \cref{Eq:R_definition}, is
\begin{align}
    \nabla \cdot \RVec
    &= \partial_n R_n
    = \partial_n
    \left\{
    \frac{1}{2} \, x_n \, \psi\psi
    + \frac{1}{2 \lambda} \,
    \Big[
    \left(D - 2 \right) \psi \, \partial_n \psi
    - x_n \, \partial_m \psi \, \partial_m \psi
    + 2 x_m \, \partial_{m} \psi \, \partial_n \psi
    \Big]
    \right\}
    \nonumber \\
    &=
    \frac{D}{2} \, \psi\psi
    + x_n \, \psi \, \partial_n \psi
    + \frac{1}{2 \lambda} \,
    \Big[
    \left(D - 2 \right) (\partial_n \psi \, \partial_n \psi + \psi \, \partial_{nn} \psi)
    - D \, \partial_m \psi \, \partial_m \psi
    - 2 x_n \, \partial_m \psi \, \partial_{mn} \psi
    \nonumber \\
    &\hspace*{4.3cm}
    + 2 \delta_{nm} \, \partial_{m} \psi \, \partial_n \psi
    + 2 x_m \, \partial_{nm} \psi \, \partial_n \psi
    + 2 x_m \, \partial_{m} \psi \, \partial_{nn} \psi
    \Big]
    \nonumber \\
    &=
    \frac{D}{2} \, \psi\psi
    + \psi \, (\vecr \cdot \nabla \psi)
    + \frac{\Delta \psi}{2 \lambda} \,
    \Big[
    \left( D - 2 \right) \psi
    + 2 \vecr \cdot \nabla \psi
    \Big] \, .
\end{align}
This divergence simplifies depending on which eigenvalue equation is satisfied by $\psi$,
\begin{align}
    -\Delta\psi & = \lambda \psi
    \quad \Rightarrow \quad
    \nabla \cdot \RVec
    = \psi^2
    \label{Eq:appendix_R_divergence}
    \, ,
    \\
    \left[-\Delta + V(\vecr) \right] \psi & = \lambda \psi
    \quad \Rightarrow \quad
    \nabla \cdot \RVec
    = \psi^2
    + \frac{V \psi}{2 \lambda} \Big[\left(D - 2 \right) \psi + 2 \vecr \cdot \nabla \psi \Big]
    \label{Eq:appendix_R_divergence_potential}
    \, .
\end{align}

The divergence of $\RVecTilde(\vecr)$, \cref{Eq:R_tilde_definition}, is
\begin{align}
    \nabla \cdot \RVecTilde
    &= \partial_n \tilde R_n
    = \partial_n
    \left\{
    \frac{1}{2} \, x_n \, \overline{\psi} \psi
    + \frac{1}{4 \lambda} \,
    \Big[
    (D - 2) \overline{\psi} \, \partial_n \psi
    - x_n \, \partial_m \overline{\psi} \, \partial_m \psi
    + 2 x_m \, \partial_m \overline{\psi} \, \partial_n \psi
    + \text{comp.\ conj.}
    \Big]
    \right\}
    \nonumber \\
    &=
    \frac{D}{2} \, \overline{\psi} \psi
    + \frac{1}{2} \, x_n \, (\overline{\psi} \, \partial_n \psi + \psi \, \partial_n \overline{\psi})
    \nonumber \\
    &\hspace*{0.4cm}
    + \frac{1}{4 \lambda} \,
    \Big[
    (D - 2) (\partial_n \overline{\psi} \, \partial_n \psi
                   +\overline{\psi} \, \partial_{nn} \psi)
    - D \, \partial_m \overline{\psi} \, \partial_m \psi
    - x_n \, (\partial_{mn} \overline{\psi} \, \partial_m \psi + \partial_m \overline{\psi} \, \partial_{mn} \psi)
    \nonumber \\
    &\hspace*{1.5cm}
    + 2 \delta_{mn} \, \partial_m \overline{\psi} \, \partial_n \psi
    + 2 x_m \, \partial_{nm} \overline{\psi} \, \partial_n \psi
    + 2 x_m \, \partial_m \overline{\psi} \, \partial_{nn} \psi
    + \text{comp.\ conj.}
    \Big]
    \nonumber \\
    &=
    \frac{D}{2} \, \overline{\psi} \psi
    + \frac{1}{2} \, \left[\, \overline{\psi} \, (\vecr \cdot \nabla) \, \psi + \psi \, (\vecr \cdot \nabla) \, \overline{\psi} \, \right]
    + \frac{1}{4 \lambda} \,
    \Big[
    (D - 2) \overline{\psi} \, \Delta \psi
    + 2  (\vecr \cdot \nabla \overline{\psi}) \, \Delta\psi
    + \text{comp.\ conj.}
    \Big]
    \, ,
    \label{Eq:appendix_R_tilde_divergence}
\end{align}
where in the last step some terms cancel when considering the complex conjugated terms.
This divergence simplifies for solutions $\psi$ to the eigenvalue equation
$-\Delta\psi = \lambda \psi$,
\begin{align}
    \nabla \cdot \RVecTilde
    &=
    \frac{D}{2} \, \overline{\psi} \psi
    + \frac{1}{2} \, \left[\, \overline{\psi} \, (\vecr \cdot \nabla) \, \psi + \psi \, (\vecr \cdot \nabla) \, \overline{\psi} \, \right]
    - \frac{1}{2} \,
    \Big[
    (D - 2) \overline{\psi} \, \psi
    + (\vecr \cdot \nabla \overline{\psi}) \, \psi
    + (\vecr \cdot \nabla \psi) \, \overline{\psi}
    \Big]
    \nonumber \\
    &= |\psi|^2
    \, ,
\end{align}
where in the first step it is used that $\lambda \in \mathbb{R}$, as otherwise
the complex conjugated terms from
\cref{Eq:appendix_R_tilde_divergence}
give rise to factors $\frac{\overline{\lambda}}{\lambda} \neq 1$.

\section{Evaluation of the asymptotic radial integral}%
\label{sec:appendix_radial_integral}

\begin{figure}[t!]
    \centering
    \includegraphics[width=0.7\columnwidth]{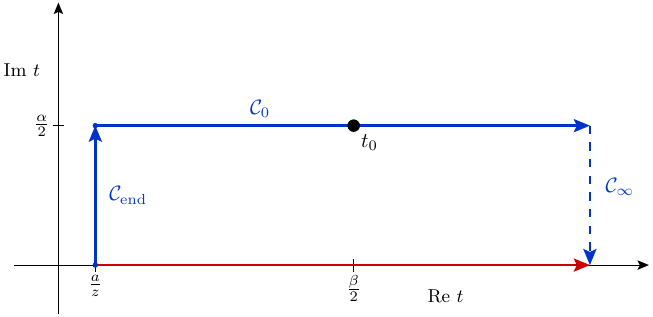}
    \caption{
        Deformation of the integration
        contour~\cref{Eq:appendix_standard_form} into the complex plane.
        The original path along the real axis (red) is deformed (blue) into
        the segments $\mathcal{C}_{\text{end}}$, $\mathcal{C}_0$, and
        $\mathcal{C}_\infty$, passing through the saddle point
        $t_0 = \frac{\beta}{2} + \ui \frac{\alpha}{2}$ (black dot) along the
        steepest-descent direction $y = \frac{\alpha}{2}$.
        }
    \label{fig:appendix_contour}
\end{figure}

In this appendix we show that the radial integral occurring in the evaluation
of the norm and biorthogonality, see the discussion
after~\cref{Eq:integral_domain_decomposition}, vanishes.
The integral is of the form
\begin{equation}
    I(a, s, \tilde{k}) = \lim_{\eta\to 0^{+}} \eta \int_{a}^{\infty} \ud r \;
    r^{s} \, \ue^{\ui \tilde{k} r} \, \ue^{-\eta r^{2}}
    \label{Eq:appendix_radial_integral}
\end{equation}
with lower bound $a > 0$, $s \in \mathbb{R}$,
$\tilde{k} = \alpha - \ui \beta \in \mathbb{C}$,
$\alpha = \Real \, \tilde{k} > 0$,
$\beta = - \Imag \, \tilde{k} > 0$, and
\begin{equation}
    \alpha > \beta \, .
    \label{Eq:appendix_condition}
\end{equation}
We evaluate \cref{Eq:appendix_radial_integral}
for $\eta \to 0^{+}$ by the
method of steepest descent.
Substituting $t = \eta r$ and setting $z = \frac{1}{\eta}$ yields the standard
form
\cite[\href{https://dlmf.nist.gov/2.4.iv}{\S2.4(iv)}]{DLMF}
\begin{equation}
    I(a, s, \tilde{k}) = \lim_{z\to\infty} z^{s} \int_{\frac{a}{z}}^{\infty}
    \ud t \; q(t) \, \ue^{- z\, p(t)} \, ,
    \label{Eq:appendix_standard_form}
\end{equation}
with $q(t) = t^{s}$ and $p(t) = t^{2} - \ui \tilde{k} t$.
The saddle point of $p(t)$ lies at
$t_0 = \ui \frac{\tilde{k}}{2} = \frac{\beta}{2} + \ui \frac{\alpha}{2}$ with
$p(t_0) = \frac{\tilde{k}^{2}}{4}$.
The path of steepest-descent through $t_0$ is the one that fulfills
$\Imag\, p(t) = \Imag\, p(t_0) = - \frac{\alpha \beta}{2}$ such that
$\ue^{-z p(t)}$ decays most rapidly.
This path is given by the line $y = \frac{\alpha}{2}$.
We now use that $p(t)$ is an entire function and restrict $q(t) = t^s$
to the principal branch.
The only singularity is at the branch point $t = 0$ which lies to the left of
the integration contour as $a > 0$.
This allows for applying Cauchy's integral theorem and deforming the
integration path onto the line $y = \frac{\alpha}{2}$ through $t_0$,
see~\cref{fig:appendix_contour}.
The integral in~\cref{Eq:appendix_standard_form}
is thus decomposed into three contributions
\begin{equation}
    I(a, s, \tilde{k}) = I_{\text{end}} + I_{0} + I_{\infty} \, ,
    \label{Eq:appendix_decomposition}
\end{equation}
which are evaluated along the vertical segment $\mathcal{C}_{\text{end}}$, the
segment $\mathcal{C}_0$ parallel to the real
axis through $t_0$, and the segment $\mathcal{C}_\infty$ closing the contour
at infinity.

\begin{itemize}
\item For the endpoint segment $\mathcal{C}_{\text{end}}$ from
$t = \frac{a}{z}$ to $t = \frac{a}{z} + \ui \frac{\alpha}{2}$ we substitute
$t = \frac{a + \ui v}{z}$ with $0 \le v \le  \frac{\alpha z}{2}$, use
$\tilde{k} = \alpha - \ui \beta$ and obtain
\begin{equation}
    I_{\text{end}} = \lim_{z\to\infty}
    \frac{\ui \, \ue^{a \beta - \frac{a^{2}}{z} + \ui a \alpha}}{z}
    \int_{0}^{\frac{\alpha z}{2}} \ud v
    \left( a + \ui v \right)^{s}
    \ue^{\frac{v^{2}}{z} - \alpha v +
    \ui \left( \beta - \frac{2a}{z} \right) v} \, .
    \label{Eq:appendix_end_integral}
\end{equation}
We now take the absolute value and apply the triangle inequality for
the integral.
The remaining integral can be bounded by a finite $z$-independent constant
$A_s = \int_{0}^{\infty} \ud v \; \left| a + \ui v \right|^{s}
\ue^{- \frac{\alpha v}{2}} < \infty$.
This is done by extending the upper integration bound to infinity and using
the negative upper bound $\frac{v^2}{z} - \alpha v \leq - \frac{\alpha v}{2}$
(the latter follows from $v \leq \frac{\alpha z}{2}$ and $v \ge 0$, giving
$\frac{v^{2}}{z} \leq \frac{\alpha v}{2}$ on the integration interval).
This yields
\begin{equation}
    \left| I_{\text{end}} \right|
    \le \lim_{z \to \infty} \frac{\ue^{a \beta - \frac{a^{2}}{z}}}{z} \, A_s
    = 0 \, ,
    \label{Eq:appendix_end_result}
\end{equation}
showing that the contribution from the endpoint segment vanishes.
\item For the saddle segment $\mathcal{C}_0$ which starts at
$t = \frac{a}{z} + \ui \frac{\alpha}{2}$ and passes horizontally through the
saddle point, we set $u = t - t_0$, so that
$p(t_0 + u) = p(t_0) + u^{2}$.
Here, $u$ extends from $u_0 = \frac{a}{z} - \frac{\beta}{2}$ to $\infty$, where
$u_0 < 0$ for sufficiently large $z$.
The resulting Gaussian integral is asymptotically evaluated
by~\cite[Eq.~(\href{https://dlmf.nist.gov/2.4.E15}{2.4.15})]{DLMF},
\begin{equation}
    I_0 = \lim_{z\to\infty}
    z^{s} \ue^{-z \frac{\tilde{k}^{2}}{4}}
    \int_{u_0}^{\infty} \ud u \;
    \left( u + t_0 \right)^{s} \ue^{-z u^{2}}
    = \lim_{z\to\infty} \sqrt{\pi} \, t_0^{s} \, z^{s - \frac{1}{2}} \,
    \ue^{-z \frac{\tilde{k}^{2}}{4}} = 0 \, ,
    \label{Eq:appendix_saddle_result}
\end{equation}
where in the last step we use that the exponential term vanishes for
$z \to \infty$ with $\Real \, \tilde{k}^{2} = \alpha^2 - \beta^2 > 0$,
following from~\cref{Eq:appendix_condition}.
We have thus shown that the contribution from the saddle segment vanishes.
\item For the segment $\mathcal{C}_\infty$ we use the vertical closing
segment $\mathcal{C}_R$ from $R + \ui \frac{\alpha}{2}$ to $R$ and take
$R \to \infty$.
For $t = R + \ui y$ with $0 \le y \le \frac{\alpha}{2}$ one has
$\Real\, p(t) = R^{2} - \beta R + \alpha y - y^{2} \ge \frac{R^{2}}{2}$ for
$R \ge 2\beta$.
Further, we have $|q(t)| \le B\, R^{|s|}$ with a finite constant $B$ for
sufficiently large $R$.
Using the triangle inequality for integrals, we thus find
\begin{equation}
    |I_{\infty}| = \lim_{z \to \infty} \lim_{R \to \infty} z^{s} \left|
    \int_{\mathcal{C}_R} \ud t \;
    q(t) \, \ue^{-z p(t)} \right|
    \leq \lim_{z \to \infty} \lim_{R \to \infty} z^{s} \, \frac{\alpha}{2} \,
    B \, R^{|s|} \,
    \ue^{-z \frac{R^{2}}{2}}
    = 0 \, ,
    \label{Eq:appendix_infinity_result}
\end{equation}
such that the contribution from infinity vanishes.
\end{itemize}
All three contributions in~\cref{Eq:appendix_decomposition}
vanish,
$I_{\text{end}} = I_{0} = I_{\infty} = 0$, and therefore
\begin{equation}
    I(a, s, \tilde{k}) = 0 \, .
    \label{Eq:appendix_radial_integral_result}
\end{equation}

\section{Electromagnetic system with a smooth boundary}%
\label{sec:appendix_smooth_boundary}

In this appendix we show that the boundary integral for the norm of
electromagnetic systems with non-dispersive, isotropic, and homogeneous
material,~\cref{Eq:norm_electrodynamics_boundary}, simplifies for a smooth
boundary instead of a piecewise-smooth boundary.
This simplification was suggested by the large language model Claude Opus 5.5.

First, we consider the term $(\vecr \cdot \nabla_{\tVec}) \nuVec$ in
\cref{Eq:norm_electrodynamics_boundary}.
Using the gradient of a dot product, one has
\begin{align}
    \nabla (\nuVec \cdot \vecr)
    &=
    (\nuVec \cdot \nabla) \vecr
    + (\vecr \cdot \nabla) \nuVec
    + \nuVec \times (\nabla \times \vecr)
    + \vecr \times (\nabla \times \nuVec)
    \nonumber \\
    &=
    \partial_\nu \vecr
    + (\vecr \cdot \nabla_{\tVec}) \nuVec
    \, ,
\end{align}
where we use the decomposition of the gradient,
\cref{Eq:nabla_normal_tangential},
$\nabla = \nuVec \partial_\nu + \nabla_{\tVec}$, in the first two terms,
$\partial_\nu \nuVec = 0$,
$\nabla \times \vecr = 0$, and $\nabla \times \nuVec = 0$.
Taking the tangential component of the above equation yields,
\begin{align}
    \nabla_{\tVec} (\nuVec \cdot \vecr) =
    (\vecr \cdot \nabla_{\tVec}) \nuVec
    \, .
\end{align}
For one of the terms in the last line of
\cref{Eq:norm_electrodynamics_boundary} we thus have
\begin{align}
    \oint_{\partial \Gamma} \dS \, H^\text{in}_\nu \EVec_{\tVec} \cdot
    [\nuVec \times (\vecr \cdot \nabla_{\tVec}) \nuVec]
    &=
    \oint_{\partial \Gamma} \dS \, H^\text{in}_\nu \EVec_{\tVec} \cdot
    [\nuVec \times \nabla_{\tVec} (\nuVec \cdot \vecr)]
    \nonumber \\
    &=
    \oint_{\partial \Gamma} \dS \, H^\text{in}_\nu [\nabla_{\tVec} (\nuVec \cdot \vecr)]
    \cdot (\EVec_{\tVec} \times \nuVec)
    \nonumber \\
    &=
    \oint_{\partial \Gamma} \dS \, \nabla_{\tVec} \cdot
    \big[ H^\text{in}_\nu (\nuVec \cdot \vecr) (\EVec_{\tVec} \times \nuVec) \big]
    - \oint_{\partial \Gamma} \dS \, (\nuVec \cdot \vecr) (\EVec_{\tVec} \times \nuVec)
    \cdot \nabla_{\tVec} H^\text{in}_\nu
    \nonumber \\
    &\hphantom{=}
    - \oint_{\partial \Gamma} \dS \, H^\text{in}_\nu (\nuVec \cdot \vecr)
    \nabla_{\tVec} \cdot (\EVec_{\tVec} \times \nuVec)
    \, ,
    \label{Eq:appendix_smooth_boundary_1}
\end{align}
where we use the cyclic property of the scalar triple product in the second step
and the product rule for the tangential divergence in the third step.
We now discuss why the first integral after the last step of
\cref{Eq:appendix_smooth_boundary_1} vanishes.
The vector field
$H^{\, \text{in}}_\nu (\nuVec \cdot \vecr) (\EVec_{\tVec} \times \nuVec)$
in \cref{Eq:appendix_smooth_boundary_1} is tangential to $\partial \Gamma$.
Together with the tangential gradient, the integrand is a divergence on the
surface $\partial \Gamma$.
For a smooth boundary $\partial \Gamma$ the vector field is continuously
differentiable on the closed surface
$\partial \Gamma$, such that the integral vanishes
by the divergence theorem in two dimensions.
We mention as an aside that for a piecewise-smooth boundary, line integrals
along the edges would remain.

Applying the cyclic property of the scalar triple product to the second term and
the divergence of a cross product to the third term of
\cref{Eq:appendix_smooth_boundary_1} gives
\begin{align}
    \oint_{\partial \Gamma} \dS \, H^\text{in}_\nu \EVec_{\tVec} \cdot
    [\nuVec \times (\vecr \cdot \nabla_{\tVec}) \nuVec]
    &=
    \oint_{\partial \Gamma} \dS \, (\nuVec \cdot \vecr) \nuVec \cdot
    (\EVec_{\tVec} \times \nabla_{\tVec} H^\text{in}_\nu)
    \nonumber \\
    &\hphantom{=}
    - \oint_{\partial \Gamma} \dS \, H^\text{in}_\nu (\nuVec \cdot \vecr)
    \left[
        \nuVec \cdot (\nabla_{\tVec} \times \EVec_{\tVec})
        - \EVec_{\tVec} \cdot (\nabla_{\tVec} \times \nuVec)
    \right]
    \, .
\end{align}
Finally, we use $\nabla_{\tVec} \times \nuVec = 0$ and
$\nuVec \cdot (\nabla_{\tVec} \times \EVec_{\tVec})
= \nuVec \cdot (\nabla \times \EVec^{\text{in}})
= \ui \omega \mu_0 \murel H^{\, \text{in}}_\nu$,
which follows from the Maxwell equation,
\cref{Eq:maxwell3}, on the inner side
of the boundary, giving
\begin{align}
    \oint_{\partial \Gamma} \dS \, H^\text{in}_\nu \EVec_{\tVec} \cdot
    [\nuVec \times \left( \vecr \cdot \nabla_{\tVec} \right) \nuVec]
    &=
    \oint_{\partial \Gamma} \dS \, (\nuVec \cdot \vecr) \,
    \left[
        \nuVec \cdot (\EVec_{\tVec} \times \nabla_{\tVec} H^\text{in}_\nu)
        - \ui \omega \mu_0 \murel \left( H^\text{in}_\nu \right)^2
    \right]
    \, .
\end{align}
Similarly, one can show for the other term in the last line of
\cref{Eq:norm_electrodynamics_boundary} that
\begin{align}
    \oint_{\partial \Gamma} \dS \, E^\text{in}_\nu \HVec_{\tVec} \cdot
    [\nuVec \times (\vecr \cdot \nabla_{\tVec}) \nuVec]
    &=
    \oint_{\partial \Gamma} \dS \, (\nuVec \cdot \vecr) \,
    \left[
        \nuVec \cdot (\HVec_{\tVec} \times \nabla_{\tVec} E^\text{in}_\nu)
        + \ui \omega \varepsilon_0 \epsrel \left( E^\text{in}_\nu \right)^2
    \right] \, .
\end{align}
Inserting these two results into
\cref{Eq:norm_electrodynamics_boundary} gives
\cref{Eq:norm_electrodynamics_boundary_smooth}.

\newpage

%%%%%%%%%%%%%%%%%%%%%%%%%%%%%%%%%%%%%%%%%%%%%%%%%%%%%%%%%%%%%%%%%%%%%%%%%%%%%

\end{document}